\documentclass{llncs}
\usepackage{algorithm}
\usepackage{algpseudocode}
\usepackage{makeidx}  
\usepackage{multirow,pifont}
\usepackage[dvips]{graphicx}
\usepackage{url}
\usepackage{flushend}
\usepackage{wrapfig}

\newcommand{\techreport}[2]{#2}

\newcommand{\codeE}[1]{\ensuremath{\tt #1}}
\def\false{\code{false}}
\def\true{\code{true}}

\newcommand{\contra}{inconsistency}
\newcommand{\toolname}{\sc{HIPCAP}} 
\newcommand{\hip}{{\sc{HIP}}} 
\newcommand{\parahip}{{\sc{ParaHIP}}} 
\def\homedir{\Large{${}_\char126$}}
\newcommand{\hipname}{\tt hipcap}
\newcommand{\toolurl}{\tt http://loris-7.ddns.comp.nus.edu.sg/{\homedir}project/{\hipname}/} 

\newcommand{\khanh}[1]{ \textit{ {\color{red} Khanh says: #1}}}

\newcommand{\todone}[1]{}
\newcommand{\ttodo}[1]{}

\newcommand{\done}[1]{}
\newcommand{\hide}[1]{}
\newcommand{\savespace}{\vspace{-2mm}}
\newcommand{\savesmallspace}{\vspace{-1mm}}

\renewcommand{\algorithmicrequire}{\textbf{Input:}}
\renewcommand{\algorithmicensure}{\textbf{Output:}}

\usepackage{amsmath,amssymb,bm,mathtools}

\usepackage{relsize} 

\usepackage{multicol}

\usepackage{frameit,color,latexsym,graphics,wrapfig}
\usepackage{listings}
\usepackage{multirow}
\usepackage{lineno}
\newcommand{\cc}[1]{{\small$\ensuremath{\tt #1}$}} 

\newcommand{\addperm}[2]{{#1}\mbox{$_{#2}$}}
\usepackage{graphicx}

\usepackage{xspace}

\usepackage{framed,color} 

\usepackage{tikz}
\usetikzlibrary{shapes,positioning,arrows,calc,fit}

\newcommand{\rpreds}[2]{\ensuremath{\btt{#1}({#2})}}
\newcommand{\rnpredl}[3]{\ensuremath{\btt{#1}({#3})}} 

\newcommand{\vlist}[1]{\ensuremath{\bar{\code{#1}}}}
\newcommand{\RS}{\ensuremath{\code{RS}}}

\newcommand{\CNT}[2]{\ensuremath{\code{CNT}(\code{#1},\code{#2})}}
\newcommand{\LatchIn}[2]{\ensuremath{\code{LatchIn}(\code{#1},\code{#2})}}
\newcommand{\LatchOut}[2]{\ensuremath{\code{LatchOut}(\code{#1},\code{#2})}}

\newcommand{\hidefa}[1]{}

\newcommand{\DEC}{\ensuremath{\code{DEC}}}
\newcommand{\KILLR}{\ensuremath{\code{KILL}}}

\def\sep{\code{*}}
\newcommand{\septract}{\ensuremath{\code{-\!-\!}\circledast}}

\newcommand{\sm}[1]{\mbox{$#1$}}
\newcommand{\btt}[1]{{\ensuremath{\tt #1}}}
\newcommand{\emp}{\btt{emp}}

\newcommand\nonterm[1]{\textit{#1}}
\newcommand\term[1]{\textit{#1}}
\newcommand\lit[1]{{\bf #1}}

\newcommand{\constr}{\ensuremath{\Phi}}

\def\pre{\constr_{\myit{pr}}}
\def\post{\constr_{\myit{po}}}

\newcommand{\requires}{\ensuremath{{\code{requires}}}}
\newcommand{\ensures}{\ensuremath{{\code{ensures}}}}
\newcommand{\res}{\ensuremath{\code{res}}}

\def\D{\Delta}

\newcommand{\myit}[1]{\textit{#1}}

\newcommand{\self}{\btt{self}}
\newcommand{\veq}{\ensuremath{\equiv}}
\newcommand{\pure}{\ensuremath{\pi}}
\newcommand{\heap}{\ensuremath{\kappa}}

\newcommand{\aheap}{\ensuremath{\iota}} 
\newcommand{\bheap}{\ensuremath{\eta}} 
\newcommand{\arith}{\ensuremath{\alpha}}

\newcommand{\aterm}{\ensuremath{\alpha^t}}

\newcommand{\set}[1]{\ensuremath{\{{#1}\}}}

\newcommand{\fracpermc}{\btt{\epsilon}} 
\newcommand{\fracperm}{\btt{\epsilon}} 
\newcommand{\grammarPerm}{\btt{\xi}} 

\newcommand{\entailS}[2]{\ensuremath{#1 \vdash #2}}
\newcommand{\entailH}[3]{\ensuremath{#1 \vdash #2 \rewrite #3}}

\newcommand{\entailK}[5]{\ensuremath{#3 \vdash^{#1}_{#2}#4 \rewrite #5}}
\newcommand{\entailC}[4]{\ensuremath{#2 \vdash_{#1} #3 \rewrite #4}}
\newcommand{\entailCV}[3]{\entailC{\myit{E}}{#1}{#2}{#3}}

\newcommand{\entailHO}[3]{\entailK{}{\btt{f}}{#1}{#2}{#3}}

\newcommand{\entailAnyVHO}[4]{\entailK{\fann}{#4}{#1}{#2}{#3}}

\newcommand{\entailInVHO}[4]{\entailK{\inflow}{#4}{#1}{#2}{#3}}

\newcommand{\defs}{\stackrel{\text{\ensuremath{\texttt{def}}}}{=}}

\newcommand{\rulen}[1]{\ensuremath{{\bf \scriptstyle #1}}}
\newcommand{\entrulen}[1]{\underline{{\bf \rulen{#1}}}}

\newcommand{\normrulen}[1]{[\underline{{\bf \rulen{NORM-#1}}}]}
\newcommand{\splitrulen}[1]{[\underline{{\bf \rulen{SPLIT-#1}}}]}

\newcommand{\errrulen}[1]{[\underline{{\bf \rulen{ERR-#1}}}]}
\newcommand{\waitrulen}[1]{[\underline{{\bf \rulen{WAIT-#1}}}]}

\def\BAG{{\cal B}}
\def\bagunion{\cup}

\def\bagsubtract{{-}}

\newtheorem{mydef}{Definition}

\def\rewrite{\leadsto}

\def\pre{\ensuremath{\constr_{pr}}}
\def\post{\ensuremath{\constr_{po}}}

\def\codesmaller{\small}

\newcommand{\code}[1]{\texttt{\textup{\codesmaller #1}}}
\newcommand{\codeS}[1]{\ensuremath{\texttt{\textup{\codesmaller #1}}}}

\newcommand\mynewcommand[2]{\newcommand{#1}{#2\xspace}}
\newcommand\blind[2]{#2} 

\mynewcommand{\hA}{\code{A}} 
\mynewcommand{\hB}{\code{B}} 

\mynewcommand{\hI}{\code{I}} 

\mynewcommand{\hJ}{\code{J}}
\mynewcommand{\hO}{\code{O}}
\mynewcommand{\ho}{\code{o}}

\mynewcommand{\hX}{\code{X}} 
\mynewcommand{\hY}{\code{Y}} 
\mynewcommand{\hL}{\code{L}} 
\mynewcommand{\hM}{\code{M}} 
\mynewcommand{\hN}{\code{N}} 
\mynewcommand{\he}{\code{e}} 
\mynewcommand{\hv}{\code{v}} 
\mynewcommand{\hl}{\code{l}} 
\mynewcommand{\lroot}{\code{l}_\top} 
\mynewcommand{\lthis}{\code{l}_\smallcode{this}} 
\mynewcommand{\hx}{\code{x}} 
\mynewcommand{\hf}{\code{f}} 
\mynewcommand{\hF}{\code{F}} 
\mynewcommand{\hT}{\code{T}} 
\mynewcommand{\hU}{\code{U}} 
\mynewcommand{\hV}{\code{V}} 
\mynewcommand{\hH}{\code{H}} 
\mynewcommand{\hS}{\code{S}}
\mynewcommand{\hsub}{\code{/}} 
\mynewcommand{\hthis}{\code{this}} 
\mynewcommand{\hclass}{\code{class}}
\mynewcommand{\hreturn}{\code{return}}
\mynewcommand{\hnew}{\code{new}}
\newcommand{\gt}{\code{>}}

\mynewcommand{\gap}{~ ~ ~ ~ ~ ~}

\mynewcommand{\st}{\ensuremath{\mathrel{{\leq}}}} 

\newcommand{\eimply}[1]{\ensuremath{{\rightarrow}}} 
\newcommand{\ourimply}{\ensuremath{\,\longrightarrow\,}}

\newcommand{\dotsaying}[1]{${\cdots}$\textup{\,#1\,}${\cdots}$}
\newcommand{\dotonly}[1]{${\cdots}$#1}
\newcommand{\saying}[1]{\#~{\color{red}\textup{\,#1\,}}}
\newcommand{\fperm}[1]{\ensuremath{{\fracperm}_{#1}}}
\newcommand{\fpermN}{\ensuremath{\fracperm}}

\newcommand{\wnnay}[1]{}

\newcommand{\stnay}[1]{}

\newcommand{\dknay}[1]{}

\newcommand{\ymnay}[1]{}

\newcommand{\scsay}[1]{{{\color{green}({\bf SC:} \textit{#1}})}}
\newcommand{\scnay}[1]{}

\newcommand{\chnay}[1]{}

\definecolor{Light}{gray}{.90}
\definecolor{Dark}{gray}{.78}

\newcommand{\permsto}[3]{{\ensuremath{{#1}\,{\xmapsto{{#3}}{\!{#2}}}}}}

\newcommand{\thread}[1]{ \ensuremath{ \mathsf{ thrd{\langle} } #1 \mathsf{ {\rangle} } } }

\newcommand{\dead}[1]{\ensuremath{\code{dead(#1)}}}

\newcommand{\OM}{\ominus}

\newcommand{\OP}{\oplus}

\newcommand{\RACE}{\code{RACE-ERROR}}
\newcommand{\DEADLOCK}{\code{DEADLOCK-ERROR}}

\newcommand{\PointTo}[2]{{\ensuremath{\code{#1}{\mapsto}\code{#2}}}}
\newcommand{\GLOBAL}[2]{\framebox{\ensuremath{\code{#1}{\mapsto}\code{#2}}}}
\newcommand{\GLOBALF}[3]{\GLOBAL{#1}{#2}\ensuremath{{\wedge}{#3}}}
\newcommand{\PointToF}[3]{\PointTo{#1}{#2}\ensuremath{{\wedge}{#3}}}
\newcommand{\LOCAL}[2]{\ensuremath{\{\!|\code{#1}{\mapsto}{\code{#2}}|\!\}}}
\newcommand{\LOCALF}[3]{\LOCAL{#1}{#2}\ensuremath{{\wedge}{#3}}}

\newcommand{\WAIT}[1]{\code{WAIT\,(\{\ensuremath{#1}\})}}
\newcommand{\WAITC}[1]{\code{WAIT\,(\ensuremath{#1})}}
\newcommand{\WAITK}[1]{\code{WAIT}}

\newcommand{\ra}{\ensuremath{\rightarrow}}

\newcommand{\CDL}{\code{CountDownLatch}}

\newcommand{\CD}{\code{countDown}}

\newcommand{\await}{\code{await}}
\newcommand{\LRA}{\sm{\longrightarrow}}
\newcommand{\SPEC}[1]{{\color{blue}#1}}
\newcommand{\BSPEC}[1]{#1}
\newcommand{\CSPEC}[1]{\ensuremath{\code{\small\#~{\color{blue}#1}}}}
\newcommand{\BCOMMENT}[1]{\ensuremath{\code{\small\#~{#1}}}}

\newcommand{\fconstr}{\ensuremath{\Phi_{\code{f}}}} 
\newcommand{\fann}{\ensuremath{\delta}}
\newcommand{\action}{\ensuremath{\code{I}}}
\newcommand{\actionperm}{\ensuremath{[\action]_{\grammarPerm}}}

\newcommand{\inflow}{\ensuremath{\ominus}}
\newcommand{\outflow}{\ensuremath{\oplus}}
\newcommand{\hvardef}{\ensuremath{{\cal D}}}
\newcommand{\rsrvar}{\btt{V}} 
\newcommand{\lemmaset}{\ensuremath{{\cal L}}} 

\newcommand{\overbar}[1]{\mkern 1.5mu\overline{\mkern-1.5mu#1\mkern-1.5mu}\mkern 1.5mu}

\newcommand{\createlatch}{\ensuremath{\code{create\_latch}}}

\newcommand{\latomsymb}{\ensuremath{\code{\sm{\langle}}}}
\newcommand{\ratomsymb}{\ensuremath{\code{\sm{\rangle}}}}
\newcommand{\criticalSec}[1]{{\latomsymb}\,#1{\ratomsymb}}

\usepackage{algpascal}
\usepackage{ifthen}
\usepackage{frameit,epsfig,amsmath,color,latexsym,graphics,wrapfig}
\usepackage{amssymb,centernot,multicol,multirow,mathtools}
\usepackage{algorithm,algpseudocode}

\usepackage{balance}

\usepackage[utf8]{inputenc}
\usepackage[T1]{fontenc}
\usepackage{microtype}

\begin{document}

\title{Automated Verification of CountDownLatch}
\hide{}
\author{Wei-Ngan Chin${}^1$ \quad Ton Chanh Le${}^1$ 
    \quad Shengchao Qin${}^2$ 
}

\institute{
${}^1$National University of Singapore\\
${}^2$Teesside University
}




\maketitle              

\pagestyle{plain} 

\begin{abstract}
\hide{
In this paper, we propose ``\myit{flow-aware}'' resource predicate,
a variant of concurrent abstract predicate,
to enable expressive verification in support of \code{countDownLatch}.
As a concurrent predicate, our flow-aware variant explicitly tracks
resources that flow into and out of its shared abstraction.
We show how to use flow-aware predicates
to safely track both the \code{countDown} and 
\code{await} mechanisms, and their support for resource
exchanges.
We demonstrate the
soundness of our flow-aware resource predicates by proving
our code and specifications to be race-free,  deadlock-free
and resource-preserving (e.g. leak-free communication).
While we were originally motivated by the need to support
\code{CountDownLatch}, our proposal for flow-aware
resource predicates can also be applied to other
concurrency mechanisms such as locks and threads.
We have implemented flow-aware predicates in a tool, called {\toolname},
on top of an existing {\parahip} verifier.
Initial implementation efforts showed that
{\toolname} is more expressive,
whilst achieving comparable verification performance.
}

The \code{CountDownLatch} (CDL) is a versatile
concurrency mechanism that
was first introduced in Java 5, and is also being adopted
into C++ and C\#. 
Its usage allows one or more 
threads to exchange resources and
synchronize by waiting for some tasks to be
completed before others can proceed.
In this paper, we propose a new framework for verifying the
correctness of concurrent applications that use CDLs. 
Our framework is built on top of two existing mechanisms,
{\em concurrent abstract predicate} and {\em fictional separation
logic}, with some enhancements such as
{\em borrowed heap} and {\em thread local abstraction}.
In addition, we propose a new inconsistency
detection mechanism based on
waits-for relation to guarantee {\em deadlock freedom}.
Prior concurrency verification works 
have mostly focused
on 
{\em data-race freedom}.
As a practical proof of concept,
we have implemented this new specification and
verification mechanism for CDL 
in a new tool, called {\toolname},
on top of an existing {\hip} verifier.
We have used this new 
tool 
to successfully verify various use cases
for CDL.

\end{abstract}


\section{Introduction}
\label{sec:intro}


One of the most popular techniques for reasoning about concurrent programs
is separation logic~\cite{OHearn:CONCUR04,Reynolds:LICS02}.
Originally, separation logic was designed to verify
heap-manipulating sequential programs, with the ability to express
non-aliasing in the heap~\cite{Reynolds:LICS02}.
Separation logic was also extended to
verify shared-memory concurrent programs,
e.g. concurrent separation logic~\cite{OHearn:CONCUR04},
where ownerships of heap objects
are considered as \myit{resource},
which can be shared and transferred among concurrent threads.
Using fractional permissions \cite{Bornat:POPL05,Boyland:SAS03}, one can
express full ownerships for exclusive write accesses and
partial ownerships for concurrent read
accesses.
Ownerships of stack variables could also be considered as
resource and treated in the same way as
heap objects~\cite{Bornat:ENTCS06}.

Most existing solutions to verify the correctness of
concurrent programs have focused on simpler concurrency
primitives, such as binary semaphores
\cite{OHearn:CONCUR04}, locks \cite{Gotsman:APLAS07,Haack:APLAS08} and first-class threads
\cite{Leino:ESOP2009,Feng:ICFP2005}. 
Lately, some solutions are beginning to emerge for
more complex concurrency mechanisms, such as
barriers \cite{Hobor:LMCS2012} and
channels \cite{Dodds:POPL11} (using only grant/wait).
The former used multiple pre/post specifications
for every thread at each barrier point to capture resource
exchanges, but this requires some non-local reasoning
over each set of multiple pre/posts.
\hide{\cite{Dodds:POPL11}.} The latter relied on higher-order 
concurrent abstract predicates (HOCAP) to provide abstract predicates
that are stable in the presence of
interfering actions from concurrent threads.
However, with the unrestricted use of higher-order predicates,
some concurrent abstract predicates proposed in \cite{Dodds:POPL11}
were unsound.

Moreover, we are not aware of any formal verification
solution that have been applied to
the more versatile {\CDL}
where multiple threads could be involved in data exchanges.
One new challenge is the need to handle two different
kinds of threads namely, (i) those that perform
non-blocking countdown operations, and (ii) those
that perform blocking await operations. We resolve
them through two new mechanisms (i) distinct abstract
predicates 
can
semantically distinguish producers from consumers
for concurrency synchronization mechanisms;
(ii) {\em inconsistency lemma} 
that can 
detect deadlocks by tracking 
when one CDL command 
is to be completed before another.
On correctness guarantees, we 
show how to ensure both
{\em race-freedom} and {\em deadlock-freedom}.
Most existing solutions on concurrency verification have focused on
only race-freedom, whilst 
some solutions \cite{Leino:ESOP2010,Le:ATVA13}
have been proposed for verifying deadlock-freedom, mostly in the
context of mutex locks and channels. 

\hide{
In this paper, 
we propose to use a novel variant of concurrent abstract predicates,
called {\em flow-aware resource} HOCAP, to help model \CDL.
This solution is based on higher-order separation logic
and will use {\em flow-annotations} and {\em resource-preservation}
to more precisely capture exchanges of
resources amongst its concurrent threads.
Our proposal for flow-aware resource HOCAP
may also be applied to other forms of concurrency
mechanisms, including locks, message-passing and barrier synchronization.
However, as a focus for the present paper, we 
show how 
flow-aware predicates
can be used to ensure the correctness of both clients
and library implementation for
\code{countDownLatch}. 
We shall also highlight what concurrency
problems are being detected by our approach.
}
\hide{
Separation logic was traditionally extended to verify
concurrent programs with parallel composition \cite{OHearn:CONCUR04}.
Recent works also extended separation logic to handle
dynamically-created threads
\cite{Gotsman:APLAS07,Hobor:thesis2011,Jacobs:POPL11,Le:ATVA13,Leino:ESOP2010}.
Hobor \cite{Hobor:thesis2011}
allows threads to be dynamically created using
fork but does not support join.
Gotsman et al. \cite{Gotsman:APLAS07} use
thread handles to represent threads,
while {\sc Chalice}~\cite{Leino:ESOP2010}
uses tokens, {\sc Verifast}~\cite{Jacobs:POPL11}
uses thread permissions, and
{\sc ParaHIP}~\cite{Le:ATVA13} uses \lit{and}-conjuncts
for the same purpose.
A fork operation returns a unique
handle/token/\lit{and}-conjunct/permission
(collectively referred to as thread token)
and a join operation on a thread token
causes the joining thread (joiner) to wait for
the completion of the thread corresponding
to the token (joinee).
However, existing works
\cite{Gotsman:APLAS07,Hobor:thesis2011,Jacobs:POPL11,Le:ATVA13,Leino:ESOP2010,Jones:SEFM12}
support reasoning about threads in a limited way:
unique tokens (representing threads)
are \underline{not} allowed to be split and shared
among different threads.
As such, existing works do not fully consider
threads in a first-class manner.
}

Our paper makes the following technical contributions:\\[-4ex]
\begin{itemize}\setlength{\itemsep}{0pt}
\item 
We propose the first formal verification for \CDL.
We use two resource predicates
and one counting predicate
\hidefa{novel
variant of
} 
to soundly track
resource synchronization
for our concurrent programs
using
{\CDL}(s).
\hidefa{
\item We show how co-variant and contra-variant subsumptions
can be correctly supported by 
flow-annotations.
Such subsumptions were not previously supported by HOCAP, rendering
them less suitable for modelling \CDL.
}
\item We provide a modular solution to the
count down mechanism
by supporting 
a {\em thread-local} abstraction on top of
the usual {\em global} view on its shared counter.
While this can be viewed as an instance of
fictional separation logic \cite{esop15:fictional}, our use of thread-local
abstraction goes beyond that by 
allowing the interference effects of parallel threads
to be modularly and precisely aggregated.
\item We provide interpretations for our abstract predicates. 
This involves a novel use of {\em septraction} operator to 
capture the notion of {\em borrowed} heap.
\item We highlight two desired concurrency properties
(i) race-freedom, 
and (ii) deadlock-freedom, that can be ensured by
formally verifying
our concurrent abstract predicates for \CDL.
A novel feature is the use of {\em {\contra}
lemmas} to help detect concurrency errors, after
they occur. 
\item We provide a prototype verifier for
\CDL.
We use it to verify several applications and
a library implementation of the \CDL.
\end{itemize}

\hide{
The rest of this paper is organized as follows.
Sec~\ref{sec:example} motivates the need for
``flow-aware'' resource predicates in
support of first-class threads, while
Sec~\ref{sec:latch} shows how \code{countDownLatch} is
being modelled. Sec~\ref{sec:lock} extends the support to hierarchical locks.
Sec~\ref{sec:progspec} introduces our core programming 
and specification languages, with a focus on modeling
and reasoning with flow-aware resource predicates.
Sec~\ref{sec:approach} presents our approach
to automated verification.
Sec~\ref{sec:soundness} presents desirable properties
from
flow-aware resource predicates,
and how they may be soundly ensured
by our proof rules.
Sec~\ref{sec:implementation} presents our prototype implementation.
Sec \ref{sec:related} summarizes related work before concluding.
}
\hide{--------------------------

Resource logic such as separation logic.
Heap locations as resource.
Threads as resource.

Threads as linear resource which can be joined by only one thread.
Limitations of existing approaches.

In this paper, we ....

--------------------------

In Java and C\#, methods to suspend a thread such as stop(), suspend() are deprecated
while the method destroy() was never implemented as they are prone to deadlocks
and could leave objects in inconsistent states.

http://docs.oracle.com/javase/7/docs/api/java/lang/Thread.html.

http://msdn.microsoft.com/en-us/library/system.threading.thread(v=vs.110).aspx

In C/C++, there is no thread destroy()
http://en.cppreference.com/w/c/thread

Threads are transferred among threads (not via lock invariants)
as invariants are not changed during the program verification,
while a thread will finally change its state from running
to terminating. Hence, a thread will never belong to
any invariants.

--------------------------
}

\section{Motivation for {\CDL}}
\wnnay{This section is to be improved by Chanh}
\label{sec:moti}
\wnnay{Highlight the versatility of
countDownLatch and its problems}
The {\CDL} protocol is a synchronization mechanism that allows one or more threads to wait
until a certain number of operations are completed by other threads. 
A {\CDL} instance is initialized with a non-negative count. A call of the
{\await} primitive blocks the current thread until the count of its latch 
reaches zero. At that point, all waiting threads are released and any 
subsequent invocation of {\await} returns immediately. Note that the count
of each latch is only decreased by one, down to zero for each invocation of 
the {\CD} primitive and cannot be reset.
The {\CDL} is a versatile mechanism that can be used for some non-trivial 
synchronization patterns. 


\hide{A countdown latch is a synchronization 
mechanism that allows one or more threads to wait until a 
certain number of operations are completed by other threads.
Each \code{CountDownLatch} is initialized to some given count.
Each \code{await} primitive blocks until the current count of its latch
reaches zero (from invocations of the
\code{countDown} primitive), after which all
waiting threads are released and any subsequent
invocation of \code{await} returns immediately.
Counters of each latch cannot be reset.} 



\begin{figure}

\vspace{-2mm}
\begin{minipage}[c]{0.6\textwidth}
\centering
\code{c = \createlatch(2);}

\vspace{-12pt}
\[
\left(
\begin{tabular}{ 
  l@{\extracolsep{\fill}} || 
  l@{\extracolsep{\fill}} || 
  l@{\extracolsep{\fill}} }
\BCOMMENT{\SPEC{$T_1$}} & \BCOMMENT{\SPEC{$T_2$}} & \BCOMMENT{\SPEC{$T_3$}}\\
\BCOMMENT{\SPEC{\code{$h{:=}h_s\code{cos}\frac{\theta}{2}$;}}} & 
\BCOMMENT{\SPEC{\code{$r{:=}h_s\code{sin}\frac{\theta}{2}$;}}} & \\
\code{countDown(c);} & \code{countDown(c);} & \code{await(c);}\\
& & \BCOMMENT{\SPEC{\code{$V{:=}\frac{1}{3}\pi r^2 h$;~}}}\\
\end{tabular}\!\!\right)
\]
\end{minipage}
\begin{minipage}[c]{0.4\textwidth}
\begin{tikzpicture}[every node/.style={anchor=base,
    text height=.8em,text depth=.2em,minimum size=7mm},
    ->, >=stealth']
    
  \matrix{
    \node[] (lbl) {\code{cnt=2}};\\
    \node[fill=gray!30,draw] (cnt_2) {$2$};\\
    \node[fill=gray!20,draw] (cnt_1) {$1$};\\
    \node[fill=gray!10,draw] (cnt_0) {$0$};\\
    \node[] (cnt_end) {};\\
  };
  
    \node[left= of cnt_2] (tm) {$T_3$};
    \node[left= of cnt_1] (t1) {$T_1$};
    \node[left= of cnt_0] (t2) {$T_2$};
  
    \node[right=15pt of cnt_2] (wait) {awaiting $\ldots$};
    \node[right=15pt of cnt_1] (cont_1) {$T_1$, continue};
    \node[right=15pt of cnt_0] (cont_2) {$T_2$, continue};
    \node[right=15pt of cnt_end] (cont_m) {$T_3$, resume};
  
  \path (tm) edge node [->, scale=.58, auto] {\code{await()}} (cnt_2);
  \path (t1) edge node [->, scale=.58, auto] {\code{countDown()}} (cnt_1);
  \path (t2) edge node [->, scale=.58, auto] {\code{countDown()}} (cnt_0);
  
  \draw [->, dashed] (cnt_1) -- node [] {} (cont_1);
  \draw [->, dashed] (cnt_0) -- node [] {} (cont_2);
  \draw [->, dashed, auto] (cnt_0) |- node [below] {\code{cnt=0}} (cont_m);
\end{tikzpicture}
%
%

\end{minipage}
\vspace{-2mm}
\savespace\savespace
\caption{A use-case of 
  {\CDL}}
\label{fig:latch_wait_for_consume}
\savespace
\end{figure}
Our first use-case of {\CDL} is when a thread 
must wait for other threads to complete their work, so that all resources
needed for its computation are available. 
As a simple example, we show
a program which 
calculates the volume $V$ of a 
circular 
cone given its slant height $h_s$ and aperture $\theta$.
In this program, the height $h$ and the radius $r$ of the cone are
concurrently computed by two threads $T_1$ and $T_2$. 
When both their computations have completed, these threads separately
invoke  the {\CD} method to inform the thread $T_3$, via awaiting on 
the latch to reach zero, 
to continue its computation for $V$ based on $h$ and $r$,
as illustrated in Fig.~\ref{fig:latch_wait_for_consume}.


In the second use-case, we leverage {\CDL} to implement a copyless 
multi-cast communication pattern, where a single send is being 
awaited by multiple receivers. As an example, let us assume 
that resource \code{P{\sep}Q}\footnote{\code{P{\sep}Q} is a separation
logic formula formed by separation conjunction {\sep} \cite{Ishtiaq:POPL01,Reynolds:LICS02}.} 
is being sent by a thread, 
which are awaited by two other threads, receiving \code{P} 
and \code{Q}, respectively. This copyless multi-cast 
communication can be modeled by the following concurrent 
program with a {\CDL} initialized to 1.



\begin{center}

\code{c = \createlatch(1);}

\vspace{-12pt}
\[
{
\left(
\begin{tabular}{ l@{\extracolsep{\fill}} || 
                 l@{\extracolsep{\fill}} || 
                 l@{\extracolsep{\fill}} }
\BCOMMENT{\SPEC{send P*Q}} ~&~ \code{await(c);} ~&~ \code{await(c);}\\
\code{countDown(c);} ~&~ \BCOMMENT{\SPEC{receive P}} ~&~ \BCOMMENT{\SPEC{receive Q}}\\
\code{\ldots} ~&~ \code{\ldots} ~&~ \code{\ldots}\\
\end{tabular}\right)
}
\]
\end{center}
The same pattern can be used to coordinate the starting of several threads,
in which the {\CD} action of the first thread is a starting signal for other awaiting threads to start at the same time.

Last but not least, a barrier synchronization can be implemented using {\CDL}. 
As another example, consider two threads, that own \code{P} and \code{Q} 
respectively, but wish to exchange their respective resources at a barrier.
We may model this scenario by using one \code{countDown} immediately
followed by \code{await}, at each of the barrier point in the two threads, 
as follows:




\begin{center}

\code{c = \createlatch(2);}

\vspace{-12pt}
\[
{
\left(
\begin{tabular}{ l@{\extracolsep{\fill}}  || l@{\extracolsep{\fill}} }
\code{\ldots} ~&~\code{\ldots}\\
\BCOMMENT{\SPEC{owns P}} ~&~ \BCOMMENT{\SPEC{owns Q}}\\
\code{countDown(c);~await(c);} ~&~\code{countDown(c);~await(c);}\\
\BCOMMENT{\SPEC{owns Q}} ~&~ \BCOMMENT{\SPEC{owns P}}\\
\code{\ldots} ~&~\code{\ldots}\\
\end{tabular}\right)
}
\]

\end{center}

As seen with these examples, the communication patterns
here are non-trivial and we shall see how 
resource predicates can help us perform formal
reasoning on \CDL.

\chnay{I think we should use a real latch example, such as the one in 
\url{http://docs.oracle.com/javase/7/docs/api/java/util/concurrent/CountDownLatch.html}. Can our current system handle this example?}


%

\section{
  Concurrent Abstract Predicates for \CDL}
\label{sec:flow}

In this section, we propose
a set of concurrent abstract
predicates that
can be used to formally model the {\CDL}. Apart from 
allowing its resources to be
more precisely tracked,
another key challenge 
faced by the {\CDL} concurrency protocol
is the ability to support
thread modular reasoning for its shared counter.
\hide{
These two aspects are why prior approaches
to concurrency verification have not solved
the {\CDL} problem.
}
We introduce two {\em resource} predicates \LatchIn{c}{P}
and \LatchOut{c}{P} to help track resource \code{P} precisely, 
and one counting predicate \CNT{c}{n} to help track
countdowns. Each set of
three concurrent abstract predicates is created to model
a \CDL, as follows:

$
\begin{array}{l}
\code{\CDL~\createlatch(n)~with~P}\\
\indent \SPEC{\requires~\code{n\sm{\gt}0}} \\
\indent \SPEC{\ensures~ \LatchIn{res}{P} {\sep}\LatchOut{res}{P}{\sep}\code{CNT(res,n)}_1}; \\
\indent \SPEC{\requires~\code{n\sm{=}0}} \\
\indent \SPEC{\ensures~ \code{CNT(res,-1)}_1}; \\
\end{array}
$
\\
Note the variable \code{res} denotes the return value of the method.
We used two pre/post specifications to describe this
constructor. In case \cc{n{=}0}, the latch cannot be
used for concurrency synchronization and is simply denoted 
by a final state of $\code{\CNT{res}{-1}}_{\fpermN}$ to
denote an expired latch whose count is definitely 0. 
(We attach fractional permission {\fpermN} to
\code{CNT} to help track sharing
on the \CDL. Such a scenario
would allow the latch itself to be de-allocated, 
where desired.)
In case \cc{n{>}0},
the \hidefa{first two flow-aware}resource predicates, \LatchIn{c}{P}
and \LatchOut{c}{P}, are used to
model the inflow 
and outflow, respectively,
of some resource 
\cc{P}
that are being exchanged by the \CDL.
Specifically, the predicate \LatchIn{c}{P} shall be used 
for the {\em consumption} of resource \cc{P}
into the {\CDL} (at {\CD} call), while the predicate
\LatchOut{c}{P} shall be used to model the {\em production} of \cc{P}
from the {\CDL} (at each
{\await} call), as shown below.


$\hspace*{-4mm}
\begin{array}{l}
\code{void~countDown(\CDL~i)}\\
~~ \SPEC{\requires~\LatchIn{i}{P}{\sep}\code{P}{\sep}\code{CNT(i,n)}_{\fpermN}\sm{\wedge}\code{n{>}0}}\\
~~ \SPEC{\ensures~ \code{CNT(i,n{-}1)}_{\fpermN}};\\ 
~~ \SPEC{\requires~\code{CNT(i,-1)}_{\fpermN}}\\
~~ \SPEC{\ensures~ \code{CNT(i,-1)}_{\fpermN}};\\
\end{array}
~
\begin{array}{l}
\code{void~await(\CDL~i)}\\
~~ \SPEC{\requires~\LatchOut{i}{P}{\sep}\code{CNT(i,0)}_{\fpermN}}\\
~~ \SPEC{\ensures~ \code{P}{\sep}\code{CNT(i,-1)}_{\fpermN}};\\
~~ \SPEC{\requires~\code{CNT(i,-1)}_{\fpermN}}\\
~~ \SPEC{\ensures~ \code{CNT(i,-1)}_{\fpermN}};
\end{array}
$

The higher-order formula \cc{P} denotes the logical resource 
added by our specifications for 
{\CDL} to support race-free resource exchanges,
as the underlying specification of {\CDL} is focused 
exclusively on its countdown counter and the blocking effects
on threads from 
{\await} calls.
The predicate \CNT{c}{n} does not capture any resources,
but is used to provide an abstract view of the counter inside
\CDL. A novel feature of this \cc{CNT} predicate is
that it could be used to support both a global view and
a thread local view. 
\hide{There are two special instances:
\CNT{c}{-1} and \CNT{c}{-2}. The former denotes
a potential decrement from a non-zero (shared) counter;
the latter denotes a counter which
has reach its final state with value \cc{0}.
The remaining possibility}
Note that \cc{\CNT{c}{n}_{\fpermN}{\wedge}n{\geq}0}
gives a thread-local view of the counter 
with a value of at least \cc{n}. As a special
case, \cc{\CNT{c}{0}_\fpermN} denotes a shared counter
that is at least \cc{0},
while \cc{\CNT{c}{-1}_{\fpermN}} denotes a shared
counter that is definitely \cc{0} (i.e. the counter has reached its final state with
value \cc{0}). 
In order to support resource consumption 
by the first pre/post specification of
{\CD} method, we require \cc{\CNT{c}{n}_{\fpermN}{\wedge}n{>}0}
to ensure race-free synchronization.

The following normalization rules show how multiple \cc{CNT} instances
are combined prior to each
formal reasoning step to provide a sound view
of the shared counter. 
The first rule shows idempotence on the final state of
\CNT{c}{-1}. The second
rule combines multiple
\cc{CNT} instances, where possible.
The third rule allows resource trapped in each
latch to be released at its final state.
This rule helps preserve (or release) unconsumed resources
from each expired latch.
%
\begin{small}
\[
\begin{array}{l}
\BSPEC{\normrulen{1}:
\CNT{c}{n}_{\fperm{1}} {\sep} \CNT{c}{-1}_{\fperm{1}} {\wedge} \code{n\sm{\leq}0} {\longrightarrow} \CNT{c}{-1}_{\fperm{1}+\fperm{2}}}
\\
\BSPEC{\normrulen{2}:
\CNT{c}{n1}_{\fperm{1}} {\sep} \CNT{c}{n2}_{\fperm{2}}\, {\wedge}\, \code{n}{=}\code{n1{+}n2} \,{\wedge}\, \code{n1,n2}{\geq}\code{0} 
{\longrightarrow} \CNT{c}{n}_{\fperm{1}+\fperm{2}}}
\\
\BSPEC{\normrulen{3}:
\LatchOut{c}{P} {\sep} \CNT{c}{-1}_{\fpermN}  {\longrightarrow} \CNT{c}{-1}_{\fpermN}{\sep}\code{P}}
\end{array}
\]
\end{small}

 
\hide{
Some examples of this normalization process are:
\begin{small}
\[
\begin{array}{l}
\CNT{c}{0} {\sep}\CNT{c}{-1} \,{\longrightarrow}\,\CNT{c}{-1}
\\
\CNT{c}{-1} {\sep}\CNT{c}{-2} \,{\longrightarrow}\,\CNT{c}{-2}
\\
\CNT{c}{n} {\sep}\CNT{c}{-1} \,{\wedge}\,\code{n>0}\,{\longrightarrow}\,\CNT{c}{n-1}\,{\wedge}\, 
\code{n>0}
\\
\CNT{c}{0} {\sep}\CNT{c}{-1} \,{\longrightarrow}\,\CNT{c}{0} {\sep}\CNT{c}{-1}
\end{array}
\]
\end{small}
}%

To support distribution to multiple concurrent threads,
we provide a set of splitting 
lemmas for our abstract 
predicates, 
as follows:
\begin{small}
\[
\begin{array}{l}
\BSPEC{\splitrulen{1}:
\LatchOut{i}{P{\sep}Q} \,{\longrightarrow}\, \LatchOut{i}{P} {\sep}\LatchOut{i}{Q}}
\\
\BSPEC{\splitrulen{2}:
\LatchIn{i}{P{\sep}Q} \,{\longrightarrow}\, \LatchIn{i}{P} {\sep}\LatchIn{i}{Q}}
\\
\BSPEC{\splitrulen{3}:
\CNT{c}{n}_{\fperm{1}+\fperm{2}} \,{\wedge}\, 
\code{n1,n2}{\geq}\code{0}  \, {\wedge}\, \code{n}{=}\code{n1{+}n2} 
\,{\longrightarrow}\,
\CNT{c}{n1}_{\fperm{1}} {\sep} \CNT{c}{n2}_{\fperm{2}} }
\end{array}
\]
\end{small}

Our splitting process is 
guided by the pre-condition expected
for each concurrent thread. 
This split occurs at the start of fork/par
operations.
For example, if an
{\await} call is 
in a thread, our split will
try to ensure that a suitable thread-local state, say
\cc{\LatchOut{i}{P}{\sep}\CNT{i}{0}},
is passed to this thread.


We illustrate an example \hide{below}%
in Fig.~\ref{fig.cdl.2}
where splitting lemmas allow relevant abstract
predicates to be made available for the modular verification
of each thread.
For simplicity in presentation, we shall omit fractional permissions
for the \code{CNT} predicate in the rest of this paper.
As such permissions must be tracked precisely,
they are always 
conserved by
lemmas and pre/post specifications.

\begin{figure*}[!htb]
\savespace
\begin{center}
\begin{small}
\code{c = \createlatch(2)~with~\SPEC{P{\sep}Q};} \\
\code{\CSPEC{\LatchOut{c}{P{\sep}Q}{\sep}\LatchIn{c}{P{\sep}Q}{\sep}\CNT{c}{2}}} \\
\code{\CSPEC{\LatchOut{c}{P{\sep}Q}{\sep}\LatchIn{c}{P}{\sep}\LatchIn{c}{Q}{\sep}\CNT{c}{0}{\sep}\CNT{c}{1}{\sep}\CNT{c}{1}}} \\\vspace{-12pt}
\[
\left(
\begin{tabular}{ 
  l@{\extracolsep{\fill}} || 
  l@{\extracolsep{\fill}} || 
  l@{\extracolsep{\fill}} }
\code{\dotonly{}} & \code{\dotsaying{create P}} & \code{\dotsaying{create Q}}\\
\!\!\code{\CSPEC{\!\!\LatchOut{c}{P{\sep}Q}{\sep}\CNT{c}{0}}}
& \code{\CSPEC{\!\!P{\sep}\LatchIn{c}{P}{\sep}\CNT{c}{1}}} 
& \code{\CSPEC{\!\!Q{\sep}\LatchIn{c}{Q}{\sep}\CNT{c}{1}}} \\
\!\!\code{await(c);} & \code{countDown(c);} & \code{countDown(c);}\\
\!\!\code{\CSPEC{{P{\sep}Q{\sep}\CNT{c}{-1}}}}
& \code{\CSPEC{\CNT{c}{0}}} 
& \code{\CSPEC{\CNT{c}{0}}} \\
\code{\dotsaying{use P*Q}} & \code{\dotonly{}} & \code{\dotonly{}} \\
\end{tabular}\!\!\right)
\]
\end{small}
\end{center}
\savespace
\caption{A {\CDL} with two threads counting down}\label{fig.cdl.2}
\savespace\savespace\savespace\savespace
\end{figure*}



\hidefa{
Let us now define the syntax of our flow-aware 
predicates.
\begin{mydef}[Syntax of Flow-Aware Predicates]
A {\em flow-aware} predicate is a concurrent abstract predicate
which may carry zero or more flow-annotated resources
 \sm{\overbar{\fconstr}}.
We can denote its syntax by
\sm{\btt{R}(\code{x},\overbar{\fconstr},\vlist{v})} 
where \sm{\fconstr} is defined as:

$
\begin{array}{lll} 
  \fconstr &  {::=}  & [\fann]\,\code{V} \mid \cdots  \qquad \text{where}\quad  
  \fann ~{::=} ~ 
  \inflow \mid \outflow 
\end{array}
$
\end{mydef}

\savespace
Take note that \sm{\overbar{\fconstr}} may be an empty list, since
each concurrent abstract predicate is an instance of
our flow-aware resource predicates. The first parameter of every predicate denotes the root
pointer to our resource predicate. Note
that \cc{\OM{V}} uses a logical variable 
\cc{V} to denote a resource that is to be
transferred into the protocol, while 
\cc{\OP{V}} denotes a resource that is to be
transferred out of the protocol.
Some of the
parameters may be un-annotated, e.g. \cc{V}.
This simply means that such parameters
are used as references, and not be considered
as resources that have flowed in and out of
some concurrency mechanism.
For example, if flow-aware predicate is being used to
model lock, a reference parameter may be used to
denote the resource invariant that will be
captured inside such a lock. 
More details on the structure of
logical formula \sm{\fconstr} are described in
Sec~\ref{sec:progspec}.
}

\hidefa{
\subsection{Ensuring Resource Preservation}
Concurrency protocols which synchronize resources between multiple
threads are expected to be leak-free. We, therefore, require
every lemma and each pre/post specification 
used to be resource-preserving.
If this was not the case, then the concurrency protocol either
leaks resources or is allowed to transform those resources that pass
through its protocols. For example, copy-oriented message-passing
channel would create new  resource to duplicate each resource
that has passed through
its send operation. Similarly, if threads 
themselves were
being captured by flow-aware abstract predicates,
we would need to support specifications that
are not resource-preserving there too.
Such protocols may thus have new
resources attributed to their pre/post specifications,
but their {\em non resource-preserving}
effects should be clearly identified.

Other than such protocols, all other concurrency
protocols (such as \CDL) are expected to be resource-preserving.
Our use of flow-annotations is {\em essential} for ensuring that each
concurrency primitive and each lemma used in our
reasoning is resource-preserving.
To formalize this
concept, we introduce the following
operator $\RS$ which is used to compute
net resource variables that are captured 
by a higher-order logic formula:
\hide{
\begin{small}
\[
\begin{array}{lll}

 \RS(\constr_1 \sep \constr_2) & \defs & \RS(\constr_1)  \bagunion \RS(\constr_2) \\

 \RS(\code{V})           & \defs & \set{\OP\code{V}} \\

 \RS(\sm{\btt{R}\hide{[\vlist{w}]}(\code{x},(\OP\code{V})::\overbar{\fconstr},\vlist{v})}) & \defs & \{\OP\code{V}\} \bagunion \RS(\sm{\btt{R}\hide{[\vlist{w}]}(\code{x}, \overbar{\fconstr},\vlist{v})}) \\
\RS(\sm{\btt{R}\hide{[\vlist{w}]}(\code{x},(\OM\code{V})::\overbar{\fconstr},\vlist{v})}) & \defs & \{\OM\code{V}\} \bagunion \RS(\sm{\btt{R}\hide{[\vlist{w}]}(\code{x}, \overbar{\fconstr},\vlist{v})}) \\
\RS(\sm{\btt{R}\hide{[\vlist{w}]}(\code{x},\code{V}::\overbar{\fconstr},\vlist{v})}) & \defs &  \RS(\sm{\btt{R}\hide{[\vlist{w}]}(\code{x}, \overbar{\fconstr},\vlist{v})}) \\



 \RS(\sm{\btt{R}\hide{[\vlist{w}]}(\code{x},[],\vlist{v})}) & \defs & \set{} \\

\RS(\constr) - \set{}             & \defs & \RS(\constr)\\


\RS(\constr_1)- (\{ \OP \code{V}\}{\cup}\BAG )        & \defs & \RS(\constr_1){\cup}\{ \OM \code{V}\} - \BAG   \\


\RS(\constr_1)- (\{ \OM \code{V}\}{\cup}\BAG )        & \defs & \RS(\constr_1){\cup}\{ \OP \code{V}\} - \BAG   \\

 
  \{ \OM \btt{V},\OP \btt{V} \} {\cup} \BAG        & \Rightarrow & \BAG  \quad \textrm{(simplification)}
\end{array}
\]
\end{small}
}%
\begin{small}
\[
\begin{array}{lll}

 \RS(\constr_1 \sep \constr_2) & {\defs} & \RS(\constr_1)  \bagunion \RS(\constr_2) \\

 \RS(\code{V})           & {\defs} & \set{\OP\code{V}} \\

 \RS(\sm{\btt{R}\hide{[\vlist{w}]}(\code{x},(\OP\code{V}){::}\overbar{\fconstr},\vlist{v})}) & {\defs} & \{\OP\code{V}\} {\bagunion} \RS(\sm{\btt{R}\hide{[\vlist{w}]}(\code{x}, \overbar{\fconstr},\vlist{v})}) \\
\RS(\sm{\btt{R}\hide{[\vlist{w}]}(\code{x},(\OM\code{V}){::}\overbar{\fconstr},\vlist{v})}) & {\defs} & \{\OM\code{V}\} {\bagunion} \RS(\sm{\btt{R}\hide{[\vlist{w}]}(\code{x}, \overbar{\fconstr},\vlist{v})}) \\
\RS(\sm{\btt{R}\hide{[\vlist{w}]}(\code{x},\code{V}{::}\overbar{\fconstr},\vlist{v})}) & {\defs} &  \RS(\sm{\btt{R}\hide{[\vlist{w}]}(\code{x}, \overbar{\fconstr},\vlist{v})}) \\



\end{array}
~~
\begin{array}{lll}

 \RS(\sm{\btt{R}\hide{[\vlist{w}]}(\code{x},[],\vlist{v})}) & {\defs} & \set{} \\

\RS(\constr) - \set{}             & {\defs} & \RS(\constr)\\


\RS(\constr_1){-} (\{ \OP \code{V}\}{\cup}\BAG )        & {\defs} & \RS(\constr_1){\cup}\{ \OM \code{V}\} {-} \BAG   \\


\RS(\constr_1){-} (\{ \OM \code{V}\}{\cup}\BAG )        & {\defs} & \RS(\constr_1){\cup}\{ \OP \code{V}\} {-} \BAG   \\

 
  \{ \OM \btt{V},\OP \btt{V} \} {\cup} \BAG        & \Rightarrow & \BAG  ~~ \textrm{(simplification)}
\end{array}
\]
\end{small}

\scnay{In the above defn: what can the underscore (\code{\_} ) represent? One reviewer asked about this.}

A pre/post condition
of the form \code{requires} \sm{\constr_1} \code{ensures} \sm{\constr_2}
is said to be {\em resource-preserving} if
$\RS(\constr_2) {-} \RS(\constr_1) {=} \{\}$. 
As an example, the specification of
\code{await} is resource-preserving since
$\RS(\code{CNT(c,-1)*P}) {-} \RS(\code{Latch(c,$\OP$P)*CNT(c,0)})$ {=} $\codeE{\{{\OP}P,{\OM}P\} {=} \{\}}$.
Similarly, each lemma $ \constr_1 {\longrightarrow} \constr_2$ 
is said to be resource-preserving
if $\RS(\constr_2) {-} \RS(\constr_1) {=} \{\}$.
All normalization and splitting lemmas used by
{\CDL} logical rules have been
shown to be resource-preserving.
In summary, once a concurrency library specification has been
shown to be resource-preserving, it is straightforward
to ensure that application codes are also 
resource-preserving through the use of classical
separation logic. This helps guarantee that our
concurrent programs are leak-free, 
an important property for languages that do
not use automatic garbage collection such as C.

}

\subsection{Ensuring Race Freedom}
In the {\CDL} protocol, we expect all {\CD} calls 
that are executed concurrently to the {\await} calls be completed
before the latter. This is important to ensure
{\em race-freedom}, since resources that are generated
from {\CD} operations must be completed (or used), before 
they are transferred to threads that are blocked by
the {\await} calls. In order to ensure this, we require
the precondition of {\CD} method to be either
\cc{\CNT{c}{n}{\wedge}n{>}0} or \cc{\CNT{c}{-1}}, 
but never \cc{\CNT{c}{0}}.
Thus, each violation on the pre-condition of
{\CD}\footnote{This violation occurs when \cc{skip()} in the third thread
of the example in Fig.~\ref{figure.cdl.1} is replaced by \cc{\CD(c)}.}
would be signaled as a potential race problem.
Another situation where race problem can occur is when resources
that are required by {\await} calls are not being synchronized
by any {\CD} call. Such an example is illustrated in Fig.~\ref{figure.cdl.1}.

\savespace\savespace
\begin{figure*}
\begin{center}
\begin{small}
\code{c = \createlatch(1)~with~\SPEC{P{\sep}Q};} \\
\code{\CSPEC{\LatchOut{c}{P{\sep}Q}{\sep}\LatchIn{c}{P{\sep}Q}{\sep}\CNT{c}{1}}} \\
\code{\CSPEC{\LatchOut{c}{P{\sep}Q}{\sep}\LatchIn{c}{P}{\sep}\LatchIn{c}{Q}{\sep}\CNT{c}{0}{\sep}\CNT{c}{1}{\sep}\CNT{c}{0}}} \\\vspace{-12pt}
\[
\left(
\begin{tabular}{ 
  l@{\extracolsep{\fill}} || 
  l@{\extracolsep{\fill}} || 
  l@{\extracolsep{\fill}} }
\code{\dotonly{}} & \code{\dotsaying{create P}} & \code{\dotsaying{create Q}}\\
\!\!\code{\CSPEC{\!\!\LatchOut{c}{P{\sep}Q}{\sep}\CNT{c}{0}}}
&\code{\CSPEC{\!\!P{\sep}\LatchIn{c}{P}{\sep}\CNT{c}{1}}} 
& \code{\CSPEC{\!\!Q{\sep}\LatchIn{c}{Q}{\sep}\CNT{c}{0}}} \\
\code{await(c);} & \code{countDown(c);} & \code{skip();}\\
\!\!\code{\CSPEC{\!\!{P{\sep}Q{\sep}\CNT{c}{-1}}}}
& \code{\CSPEC{\!\!\CNT{c}{0}}} 
& \code{\CSPEC{\!\!Q{\sep}\LatchIn{c}{Q}{\sep}\CNT{c}{0}}} \\
\code{\dotsaying{use P*Q}} & \code{\dotonly{}} & \code{\dotonly{}} \\
\end{tabular}\!\!\right)
\]

\code{\CSPEC{{P{\sep}Q{\sep}\CNT{c}{-1}}
 \sep \CNT{c}{0} 
 \sep Q{\sep}\LatchIn{c}{Q}{\sep}\CNT{c}{0}}} \\
\code{\CSPEC{{P{\sep}Q{\sep}\CNT{c}{-1}}
 \sep Q{\sep}\LatchIn{c}{Q}}} \\
\code{\CSPEC{{\color{red} RACE-ERROR detected by \errrulen{1} \hide{when \sm{\neg(\code{Q}=\emp)}}}}}

\end{small}
\end{center}
\savespace\savespace
\caption{A {\CDL} with a race error}\label{figure.cdl.1}\savespace\savesmallspace
\end{figure*}

The third thread has the resource \cc{Q} generated
there but was not being synchronized by any {\CD} call.
As a consequence, \cc{Q} was not properly transferred to
its corresponding {\await} call.
We propose to detect such potential race violations
through the following  {\contra}  lemma:

\indent
\BSPEC{\errrulen{1}: \LatchIn{c}{P}{\sep}\code{CNT(c,-1)}\hide{\sm{\wedge}\code{n{<}0}}\hide{{\sm{\wedge\neg}(\code{P}={\emp})}}  \sm{\longrightarrow} \RACE}

The formula
\cc{\LatchIn{c}{P}{\sep}\CNT{c}{-1}}
denotes a contradiction.
The former predicate requires the shared counter to be non-zero, 
while latter predicate is in a final state with value \cc{0}
(see Sec~\ref{sec:interpret}).
Such contradictions are manifestations of 
some concurrency synchronization errors.
We identify this as a race problem.

\savespace\savespace
\subsection{Ensuring Deadlock Freedom}
\label{sec:deadlock}
Deadlock may occur when blocking operations, such
as {\await}, are invoked and then could wait forever,
e.g., due to the shared counter never reaching zero.

For a single {\CDL}, this deadlock error can be signified
by the following lemma:

\BSPEC{\errrulen{2}: 
\code{CNT(c,a)}{\sep}\code{CNT(c,-1)}\sm{\wedge}\code{a{>}0} 
\sm{\longrightarrow} \DEADLOCK}\\
Here, \cc{CNT(c,a){\wedge}a{>}0} denotes a counter value of 
at least 1, while \cc{CNT(c,-1)} denotes a final
counter with value \cc{0}. Such a contradiction is a manifestation
of a deadlock error which arose
from the unreachability on precondition
\cc{\CNT{c}{0} \vee \CNT{c}{-1}}
that we impose on the {\await} method.
%
A simple example of this deadlock scenario
(omitting \code{Latch} predicates) is shown in Fig~\ref{fig.cdl.deadlock}. 
This deadlock error occurs due to the first thread
not invoking a sufficient number of {\CD} calls.
This lack of {\CD} calls led to an error scenario when the 
conflicting states from the two threads are being joined.

\savesmallspace
\begin{wrapfigure}{l}{0.48\textwidth}
\vspace*{-9mm}
\begin{center}
\begin{small}
\code{c = \createlatch(2);}
\\
\code{\CSPEC{CNT(c,2) {\LRA} CNT(c,2){\sep}CNT(c,0)}}\\
%
\vspace{-12pt}
\[
{
\left(
\begin{tabular}{ l@{\extracolsep{\fill}}  || l@{\extracolsep{\fill}} }
\code{\CSPEC{CNT(c,2)}} &~~\code{\CSPEC{CNT(c,0)}}\\
\code{countDown(c);~~} &~~\code{await(c);}\\
\code{\CSPEC{CNT(c,1)}} &~~\code{\CSPEC{CNT(c,-1)}}\\
\end{tabular}\right)\code{;}
}
\]

\code{\CSPEC{CNT(c,1){\sep}CNT(c,-1)}}
\\
\code{\CSPEC{{\color{red}{\DEADLOCK} detected by \errrulen{2}}}}
\end{small}
\end{center}
\vspace*{-6mm}
\caption{An intra 
deadlock scenario}\label{fig.cdl.deadlock}\vspace*{-4mm}\savespace\savespace
\end{wrapfigure}
\noindent 
For multiple latches, we will need to track
a wait-for graph \cite{Silberschatz:OS13},
to help detect deadlocks, where possible.
Let us introduce a
\addperm{\WAITC{\codeE{S}}}{\fracperm} relation that is tracked inter-procedurally
with its wait-for graph \codeE{S} and its permission \fracperm.
Whenever we have a complete view of the
waits-for relation (with full permission, denoted as \addperm{\WAITC{\codeE{S}}}{1}),
we can always
reset each acyclic wait-for graph to \code{\{\}},
as follows: \BSPEC{\waitrulen{1}: \addperm{\WAITC{\codeE{S}}}{1}\sm{{\wedge}\neg}\code{isCyclic(S)}\sm{\longrightarrow} \addperm{\WAITC{\{\}}}{1} }

\noindent Note that \code{isCyclic(S)} returns \code{\true} when the wait-for graph \code{S}
contains a cycle.  We add a wait-for arc via this lemma:

\BSPEC{\waitrulen{2}: 
\codeE{CNT(c_1,a)}{\sep}\codeE{CNT(c_2,-1)}{\sep} \addperm{\WAITC{\code{S}}}{\fracperm}\sm{\wedge}\code{a{>}0} 
\sm{\longrightarrow}}

\BSPEC{\qquad\qquad\qquad\codeE{CNT(c_1,a)}{\sep}\codeE{CNT(c_2,-1)}\sm{\wedge}\code{a{>}0} {\sep} \addperm{\WAITC{\code{S}{\cup}\codeE{\{c_2{\ra}c_1\}}}}{\fracperm}}

\noindent
An 
arc \codeE{c_2{\ra}c_1} is added into the \code{WAIT} relation
to indicate that (the counting down on) \codeE{c_2} will be completed
before the \codeE{\CDL~c_1} or in other words, \codeE{c_1} is waiting 
for \codeE{c_2} to complete. 
The wait-for predicates are split/normalized by:

\BSPEC{\waitrulen{3}: \addperm{\WAITC{\codeE{S_1}}}{\fracperm_1} {\sep} \addperm{\WAITC{\codeE{S_2}}}{\fracperm_2}  \sm{\longleftrightarrow} \addperm{\WAITC{\codeE{S_1 {\cup} S_2}}}}{{\fracperm_1}{+}{\fracperm_2}}



We normalize where possible, and split the \code{WAIT} relation
at fork/par locations.
Moreover, any occurrence of a cycle in the wait-for graph is immediately
detected as a potential deadlock by the following {\contra} lemma:

\BSPEC{\errrulen{3}: \addperm{\WAITC{
\code{S}
}}{\fracperm} 
\code{\sm{\wedge}~isCyclic(S)}
\sm{\longrightarrow} \DEADLOCK }
\\
An example of deadlock detection for multiple latches 
is shown in Fig~\ref{fig.deadlock.waitfor} 
where the deadlock is detected by the cycle 
in \codeE{\WAIT{\codeE{c2{\ra}c1, c1{\ra}c2}}}.







\savespace\savespace
\begin{wrapfigure}{l}{0.65\textwidth}
\begin{small}
\begin{center}
\code{c1 = \createlatch(1); c2 = \createlatch(1);}

\code{\CSPEC{\addperm{\WAIT{}}{1}{\sep}CNT(c1,1) {\sep} CNT(c2,1)  {\LRA}}}

\code{\CSPEC{CNT(c1,1) {\sep} CNT(c2,0) {\sep} CNT(c2,1) {\sep} CNT(c1,0) }}


\vspace{-12pt}
\[
{
\left(
\begin{tabular}{ l@{\extracolsep{\fill}}  || l@{\extracolsep{\fill}} }
\code{\CSPEC{CNT(c1,1){\sep}CNT(c2,0)}} &~~\code{\CSPEC{CNT(c2,1){\sep}CNT(c1,0)}}\\
\code{await(c2);~~} &~~\code{await(c1);}\\
\code{\CSPEC{CNT(c1,1){\sep}CNT(c2,-1)}} &~~\code{\CSPEC{CNT(c2,1){\sep}CNT(c1,-1)}}\\
\code{\CSPEC{\quad\sep\addperm{\WAIT{\codeE{c2{\ra}c1}}}{\fracperm_1}}}  &~~\code{\CSPEC{\quad\sep\addperm{\WAIT{\codeE{c1{\ra}c2}}}{\fracperm_2}}}  \\
\code{countDown(c1);~~} &~~\code{countDown(c2);}\\
\code{\CSPEC{CNT(c1,0){\sep}CNT(c2,-1)}} &~~\code{\CSPEC{CNT(c2,0){\sep}CNT(c1,-1)}}\\
\code{\CSPEC{\quad\sep\addperm{\WAIT{\codeE{c2{\ra}c1}}}{\fracperm_1}}}  &~~\code{\CSPEC{\quad\sep\addperm{\WAIT{\codeE{c1{\ra}c2}}}{\fracperm_2}}}  \\
\end{tabular}\right)
}
\]

\code{\CSPEC{\quad CNT(c1,-1){\sep}CNT(c2,-1){\sep}\addperm{\WAIT{\codeE{c2{\ra}c1,c1{\ra}c2}}}{1} }} \\
\code{\CSPEC{{\color{red}{\DEADLOCK} detected by \errrulen{3} }}}
\end{center}
\savespace\savespace\savesmallspace
\caption{An example of inter-latch deadlock 
}\label{fig.deadlock.waitfor}
\end{small}
\vspace{-1.2cm}
\end{wrapfigure}

Though deadlock detection for locks and channels have been
proposed before \cite{Leino:ESOP2010}, there are at least two novel ideas
in our 
proposal. Firstly, we have now shown how 
to formally ensure deadlock-freedom for the more complex
{\CDL}
protocol. Secondly,
we have achieved this through the use of 
{\contra}
lemmas. The wait-for arcs added are used
for {\contra}
detection, since each arc \cc{c_2{\ra}c_1} 
denotes a strict completion ordering \cc{c_2{<}c_1} (for a pair
of latches) that is
to be expected from its concurrent execution.
Any cycle from  such an accumulated
waits-for ordering denotes potential unsatisfiability.




\subsection{Interpretations for Abstract Predicates}


\label{sec:interpret}
We have introduced three concurrent abstract predicates, 
namely  \LatchIn{c}{P},  \LatchOut{c}{P} and \CNT{c}{n},
for the specification of {\CDL}. The first two predicates
are concerned with managing the flows of resource \code{P} through
its \CDL. We can provide the following interpretations for them:

\begin{small}
$
\begin{array}{lll}
 \LatchOut{i}{P} & \defs & \GLOBAL{i}{0} {\ourimply} \code{P}
\\
 \LatchIn{i}{P} &\defs &  (\code{P}{\septract}\emp) ~{\sep}~ [\DEC]_{\fracpermc}  {\sep}\GLOBAL{i}{m} {\wedge}\code{m>0} 
 \\
\end{array}
$
\end{small}

Each formula \code{\GLOBAL{i}{m}} denotes a shared global location,
while \code{\PointTo{i}{m}} denotes a fully-owned
heap (local) location. 
Predicate
\code{\LatchOut{i}{P}} is itself a {\em producer}
of resource \code{P} that is released once
shared global counter becomes \code{0}.
For the \code{\LatchIn{i}{P}} predicate,
we use \code{P{\septract}\emp} with 
a {\em septraction} operator \sm{\septract} (\cite{Vafeiadis:CONCUR07}) 
to capture the
{\em consumption} of resource \code{P} into the {\CDL}
via its \code{countDown(i)} method. Unlike
the septraction operator in \cite{Vafeiadis:CONCUR07}
which works with {\em real} heaps, our
formula \code{P{\septract}\emp} is an extension to
capture the concept of {\em virtual} heap 
that denotes a borrowing of heap \code{P}.
For example, \code{(\PointTo{x}{\_}){\septract}\emp} is a borrowing of 
\code{(\PointTo{x}{\_})} such that
\codeS{$\forall \code{Q} {\cdot} (\code{Q}{\septract}\emp){\sep}\code{Q} = \emp$}.
This simple (but novel) concept allows us to capture notion of resource flows 
through the \CDL.
The \code{\LatchIn{i}{P}}  predicate also captures a partial
permission for
{\DEC} action that causes
its shared global counter \GLOBAL{i}{m}
to be decreased by \cc{1}, as captured by:
\hide{
where the first two  carry
some logical resources (\code{P}), and the third one is used to
track the shared counter. A simple
implementation of this {\CDL} concurrency
protocol can be described by the following methods.


$
\begin{array}{l}
\code{{\CDL}\, {\createlatch}(int\, n) \{}
\code{return new \CDL(n); \}}
\\
\code{void\,countDown({\CDL} i) \{ }
\code{\criticalSec{if (i.val>0) i.val = i.val-1;\hide{else notifyAll();}}\ \}}
\\
\code{void await({\CDL} i) \{ while (i.val>0) skip; \}}
\end{array}
$
}
\hide{
Of the two concurrent methods, only the {\CD} method
may cause interference. Following \cite{Dinsdale-Young:ECOOP10},
this potential interference
can be described by a {\DEC} action
where the shared global counter \GLOBAL{i}{n}
is decreased by \cc{1} if it is non-zero.

}
\hide{The {\DEC} action is specified by
the following 
interference:
}

\begin{small}
$
\begin{array}{l}
\DEC: \begin{array}[t]{ll}
\GLOBAL{i}{n}\,{\wedge}\,\code{n{>}0} & \rightsquigarrow \GLOBAL{i}{n{-}1}
\end{array} \\
\end{array}
$
\end{small}
\hide{
We use this implementation to show how our concurrent abstract
predicates may be defined, and also to illustrate
how the correctness of {\CDL} may be verified.
In the {\CD} method, this action is currently wrapped
by an atomic operation, marked with \code{\criticalSec{$\cdots$}}.
}

To support local reasoning with updates,
we propose
a {\em thread-local} view for global counters
based on fictional separation logic \cite{Jensen:ESOP12}.
We introduce a new
{\em thread-local} formula \LOCAL{i}{n} to denote a
shared counter whose value is at least
\cc{n}.
Such a thread-local 
abstraction is related to its
global counter by the property:

\begin{small}
$
\LOCAL{i}{n} \longrightarrow \GLOBALF{i}{m}{\code{m}{\geq}\code{n}{\geq}\code{0}}
$
\end{small}

While the formula \code{\GLOBAL{i}{m}} provides a precise view
of some shared global location, the thread-local formula \code{\LOCAL{i}{m}$_{\fracperm}$}
provides a fictional thread-local view of the same counter
that could be separately updated by each thread.
This applies even if fractional permission is being
imposed for the thread-local predicate, as long as
such predicates can be shown to be stable.






To support concurrency,
we provide 
combine/split operations that can be used
to precisely handle multiple thread-local states
from concurrent threads.

\begin{small}
$
\LOCAL{i}{n}_{\fracperm_1+\fracperm_2}{\wedge}\code{a,b}{\geq}\code{0}{\wedge}{\code{n=a+b}} ~{\longleftrightarrow}~ \LOCAL{i}{a}_{\fracperm_1}{\sep}\LOCAL{i}{b}_{\fracperm_2}
$
\end{small}

\noindent
Contrast this to the global view of a shared counter with 
trivial property:

\begin{small}
$
\GLOBAL{i}{a}{\sep}\GLOBAL{i}{b} 
~{\longleftrightarrow}~\GLOBAL{i}{a}{\wedge}\code{a{=}b}
$
\end{small}

With these two views, we provide the following 
intepretation for \code{\CNT{i}{n}}:

\begin{small}
$
\begin{array}{lll}
\CNT{i}{n} &\defs & \LOCAL{i}{n} {\wedge}\code{n${\geq}$0} ~{\vee}~ \GLOBAL{i}{0}{\wedge}\code{n=-1}
\end{array}
$
\end{small}

\hide{
\noindent which  signifies that all threads have the same view for the shared counter (the \sm{\longrightarrow} direction) and
 that the global view 
can be  duplicated and shared among threads (the \sm{\longleftarrow} direction).



With this use of thread-local abstraction, we are now ready 
to provide interpretations
for our concurrent abstract predicates that are
stable, as follows:
Using this, we can now define 
\begin{small}
$
\begin{array}{lll}
 \LatchIn{i}{P} &\defs &  \code{P}   ~{\septract}~  ([\DEC]_{\fracpermc}  {\sep}\GLOBAL{i}{m} {\wedge}\code{m>0} )
 \\
 \LatchOut{i}{P} & \defs & \GLOBAL{i}{0} {\ourimply} \code{P}
\\
\CNT{i}{n} &\defs & \LOCAL{i}{n} {\wedge}\code{n${\geq}$0} ~{\vee}~ \GLOBAL{i}{0}{\wedge}\code{n=-1}
\end{array}
$
\end{small}
\\
}

To confirm soundness, we must determine that
all three concurrent abstract predicates are stable in
the presence of interferring actions.
Our use of \cc{\LOCAL{i}{n}} remains stable since
it is based on fictional separation logic, whose effect is
not affected by other concurrent threads.
The shared
global state \cc{\GLOBAL{i}{0}} is stable,
since \cc{\DEC} operation does not modify such a final state
of the shared counter.
The condition \cc{\GLOBALF{i}{m}{m{>}0}} of the shared global state 
in \cc{\LatchIn{i}{P}} is stable since it is always 
used with \cc{\CNT{i}{n}} in the pre-condition
of {\CD} method, which results in
\cc{\GLOBALF{i}{m}{m{>}0}{\sep}\LOCALF{i}{n}{n{>}0}}.
By itself the global view \cc{\GLOBALF{i}{m}{m{>}0}} would not be
stable in the presence of concurrent \cc{\addperm{{[\DEC]}}{\fracperm}}.
However, the combined state 
\cc{\GLOBALF{i}{m}{m{>}0}{\sep}\LOCALF{i}{n}{n{>}0}} remains
stable since the thread-local view of
\cc{\LOCALF{i}{n}{n{>}0}} is not affected
by \cc{\addperm{{[\DEC]}}{\fracperm}} 
operations from 
other threads.
Thus, \cc{\GLOBALF{i}{m}{m{>}0}} always hold
since \cc{m{\geq}n}.
Lastly, \LatchOut{i}{P} is also stable since it only depends
on a stable global formula \GLOBAL{i}{0} with final value \code{0}.

\dknay{ not sure if \GLOBALF{i}{m}{m{>}0}{\sep}\LOCALF{i}{n}{n{>}0}
is indeed stable, for example:
\[
\begin{array}{l}
\GLOBALF{i}{m}{m{>}0}{\sep}\LOCALF{i}{n}{n{>}0}
\\
\Rightarrow
\GLOBALF{i}{m}{m{>}0}{\sep}
\GLOBAL{i}{m} {\wedge} m{\geq}n{>}0
\\
\Rightarrow \GLOBALF{i}{m}{m{\geq}0}
\end{array}
\]
\GLOBALF{i}{m}{m{>}0} is exactly the LHS of the action DEC,
hence it is not stable under the interference of DEC.
 }
\wnnay{Above do not hold due to thread-local
{|i->n|} that is not affected by other threads}

\hide{
For the purpose of creating and later destroying this
shared global counter, we provided another action, called \cc{\KILLR} action, 
that allows a global counter to be made local by
the following equivalence lemma:
\begin{small}
$
\begin{array}{l}
\KILLR: \begin{array}[t]{ll}
\addperm{\KILLR}{1} \,{\sep}\,\addperm{\DEC}{1} \,{\sep}\, \GLOBAL{i}{n} & \rightsquigarrow \PointTo{i}{\CDL(n)}
\\[0.5ex]
\end{array}
\end{array}
$
\end{small}
\scsay{I don't quite get the above: it looks like an action definition for KILL but KILL itself is used in it; if it's an equivalence
lemma, then maybe we should use \sm{\longleftrightarrow}? For the latter case, do we need to give the action definition of KILL?}

\noindent 
Our action is a generalization of \cite{Dinsdale-Young:ECOOP10}, as
it allows both local and shared states in 
its transitions.
This action is also supported by
the following equivalence lemma:

\begin{small}
\[
\begin{array}{ll}
\PointToF{i}{\CDL(n)}{\code{n}{\geq}\code{0}}
 & {\longleftrightarrow}~ \addperm{{[\DEC]}}{1} {\sep}
\addperm{{[\KILLR]}}{1} {\sep}  \GLOBALF{i}{n}{\code{n}{\geq}\code{0}} \\
& {\longleftrightarrow}~
\addperm{{[\DEC]}}{1} {\sep} \addperm{\LOCAL{i}{n}}{1} {\sep} \GLOBAL{i}{n}
\end{array}
\]
\end{small}

This {\KILLR} action permits the creation of
a thread-local heap state.
By tracking the permissions of both \cc{\DEC} action
and thread-local counter, 
we could recover the full permission needed for the 
thread-local counter to be converted to
just the global view, prior to its
conversion back to a local heap for
explicit disposal.
}
{\bf Comparing with Fictional Separation Logic }
Our use of thread-local abstraction \LOCAL{i}{n}$_1$
(equivalent to
\GLOBAL{i}{n}
)
is an orthogonal enhancement for 
fictional 
separation logic \cite{Jensen:ESOP12}.
Though fictional abstraction can be used to
reason separately about updates to a shared 
resource amongst concurrent threads, this
abstraction is not truly {\em thread-local} since 
its split/aggregate rule does not allow effects from 
{\em strong-updates} to be {\em precisely} propagated
amongst concurrent threads.
For example, consider the monotonic (increasing)
counter from \cite{Pilkiewicz:TLDI11}. Though fictional 
abstraction has the following split/aggregate rule to 
duplicate or merge the predicate \cc{MC(c, i)} of a 
monotonic counter \cc{c}

\begin{small}
\cc{\forall i{\leq}j \cdot 
MC(c, j)_{e1+e2} \longleftrightarrow
MC(c, j)_{e1} \sep MC(c, i)_{e2}}
\end{small}

\noindent we do not consider it to be thread-local
since each predicate in this rule may not maintain
a precise view of the shared global counter.
Therefore, any update on a predicate
via local reasoning may cause other copies to become weaker.
To make this abstraction thread-local, we
will need to utilize the following conversion:

\begin{small}
\cc{
MC(c, n)_1 {\wedge} j{\leq}n
\longleftrightarrow
\LOCAL{c}{n}_1 \sep MC'(c, j)_1}
\end{small}

\noindent Here, \code{\LOCAL{c}{n}} is our precise thread-local abstraction,
while \code{MC'(c, j)} is the monotonic counter proposed 
from 
\cite{Jensen:ESOP12}.
Adding such a thread-local abstraction permits
strong updates to be precisely tracked across
concurrent threads.

\wnnay{to move 3.4-3.6 to Appendix}

\section{Formalism of Language and Logic}
\wnnay{SC: Could you improve this section?}
\label{sec:proglang}
\vspace{-1cm}
\begin{wrapfigure}{l}{0.63\textwidth}
\savespace
\savespace
\savespace
\savespace
\begin{center}
\begin{frameit}
\savespace
\savespace
\savespace
\begin{small}
\[
\begin{array}{cll}
 \nonterm{Prog} & ::= & \overbar{\nonterm{datat}} ~~ 
\overbar{\nonterm{proc}} \\

\nonterm{datat} & ::= & {\bf data}~ \code{C} ~\{~
\overline{\nonterm{t}~\term{f}} ~\}   \\



\nonterm{proc} & ::= & \nonterm{t} ~ 	
\term{pn}(\overline{\nonterm{t}~\term{v}}) ~~ \overbar{\nonterm{spec}}
~\{~\nonterm{e}~\}   \\ 
		
 \nonterm{spec} & ::= & \lit{\requires} ~\pre ~\lit{\ensures}~
\post\lit{;}  \\


 \nonterm{t} & ::= &\btt{void} \mid \btt{int} \mid \btt{bool} \mid \hide{\btt{thrd}}\hide{ \btt{lock} \mid} \btt{\CDL} \mid \code{C} \\

\nonterm{e} & ::= & 
\nonterm{v} \mid \nonterm{v.f} \mid \term{k}  \mid \lit{\code{new}}~\code{C}\lit{(} \overline{\term{v}} \lit{)} \hide{\mid \lit{\code{destroy}}\lit{(} \term{v} \lit{)}}
\mid e_1;e_2 \mid e_1 \| e_2  \mid \criticalSec{e} \quad  \\ 
& & \lit{\createlatch} \lit{(} \nonterm{n} \lit{)} ~ \lit{\code{with}}~ {\heap{\wedge}\pure}  \mid \lit{\code{countDown}} \lit{(}\nonterm{v} \lit{)}
        \mid   \\ 
& & \lit{\code{await}} \lit{(} \term{v} \lit{)} \mid   \nonterm{pn} \lit{(} \overline{\term{v}} \lit{)}  \mid \code{if} ~ \nonterm{v}~\code{then}~\nonterm{e}_1 ~\code{else}~ \nonterm{e}_2 \mid   \ldots \\

\hide{ & &  \lit{\code{create\_thread}} \lit{(} \nonterm{pn} \lit{)} \mid \lit{\code{fork}} \lit{(}\nonterm{v}, \overline{\nonterm{v}} \lit{)}
        \mid \lit{\code{join}} \lit{(} \term{v} \lit{)} \mid & \\ }

\hide{ & &  \lit{\code{create\_lock}} \lit{(}  \lit{)}~\code{with}~\constr \mid \lit{\code{acquire}} \lit{(}\term{l} \lit{)}
        \mid \lit{\code{release}} \lit{(} \term{l} \lit{)}   \mid   \lit{\code{dispose}} \lit{(} \term{l} \lit{)} & \\}

	



			

\end{array}
\]
\end{small}
\savespace
\end{frameit}
\savespace
\savesmallspace
\caption{Core Language with \hide{Threads and }{\CDL}\hide{ and Locks}}
\label{fig:proglang}
\savespace
\savespace
\savespace
\savespace
\end{center}
\end{wrapfigure}

\wnnay{I think comments of Fig 1 
should be in LHS similar to Fig 2 for logic}


We use the core language in Fig.~\ref{fig:proglang}
to formalise our reasoning with 
\hide{flow-aware resource predicates.
In this paper, we restrict our attention to}
{\CDL}.\hide{, we will add support for first-class threads and locks
in the following sections.}
\chnay{We could link these features to a real programming language, 
such as Java.}
A program consists of
data declarations ($\overbar{\nonterm{datat}}$),
and procedure declarations ($\overbar{\nonterm{proc}}$).
Each procedure declaration is annotated with
pairs of pre/post-conditions ($\constr_{pr}/\constr_{po}$).
\hide{A \lit{\code{fork}} receives a thread identifier \nonterm{pn} and
a list of parameters $\term{v}^*$, starts the execution
of the thread.
$\lit{\code{join}} \lit{(} \term{v} \lit{)}$ waits for the thread
that is referred to by \term{v} to finish its execution.
At run-time, the joiners wait for the joinee to complete
its execution.
We do not allow canceling a thread.}
\wnnay{describe latches, countDown and await}
A (countdown) latch, created by $\lit{\createlatch} \lit{(} \nonterm{n} \lit{)}$,
is initialized to a given count $\nonterm{n}$ and can be passed in with
a (logical) resource \sm{\heap{\wedge}\pure}. 
\chnay{\lit{\code{create\_latch}} is not a primitive method in 
\code{java.util.concurrency} libary, why we do not use \code{new 
CountDownLatch}?}
A \code{countDown({v})} operation decrements the count of latch \code{v}.
An \code{await({v})} operation blocks until the  count of  latch
\code{v} reaches zero, after which all
waiting threads are released and any subsequent
invocation of \code{await} returns immediately.
\scnay{Explain lock-related operations here?}
The operation \sm{\langle \textit{e} \rangle} denotes an atomic action.
Other program constructs are standard as can be found
in the mainstream languages.

\hide{We keep the core language concise in order to illustrate
the key concepts while still allow it to be extended.
For example, we could support return a value after join by passing
an additional variable by reference to capture the return
value.}

\label{sec:progspec}


\begin{figure}[htb]
\savespace
\begin{center}
\begin{minipage}{25pc}
\begin{frameit}
\savespace
\savespace
\begin{small}
\[
\begin{array}{rrll}



\!\!\!\!\textsf{FA~Pred.} & \nonterm{rpred} & {::=} & 
\code{pred}~\btt{R}(\self,\overbar{\rsrvar},\vlist{v}) 
~[\veq~\constr]~[\lit{inv}~\pure]\\

\!\!\!\!\textsf{Action} & \nonterm{act} & {::=} &
\code{action}~\action~\veq~ 
~\overbar{\aheap_1{\wedge}{\pure_1} \rightsquigarrow \aheap_2{\wedge}{\pure_2}
}\\ 

\!\!\!\!\textsf{Disj.~formula} & \constr & {::=} & \bigvee ( \exists \vlist{v} \cdot
 {\bheap}{\sep}
\heap {\wedge} \pure) \\

\!\!\!\!\textsf{Non-Resource}
& \!\! \bheap & {::=} & {\actionperm} \mid \WAIT{\overbar{v_1{\ra}v_2}}\addperm{}{\grammarPerm} \mid
\GLOBAL{\code{v}}{\code{C}(\vlist{v})} 
\mid \bheap_1 {\sep} \bheap_2
\\
 
\!\!\!\!\textsf{Sep.~formula} & \heap & {::=} & 
\aheap
\mid 
\hidefa{[\sm{\OM}]\,}\rsrvar
\mid \rnpredl{\btt{R}}{\vlist{v}}{\code{v},\overbar{\fconstr},\vlist{v}}
\mid  \heap_1 \sep \heap_2 \\

\!\!\!\!\textsf{Simple~heap}
& \!\! \aheap & {::=} &
\emp
\mid \permsto{\code{v}}{\code{C}(\vlist{v})}{}
\mid \LOCAL{\code{v}}{\code{C}(\vlist{v})}
\mid \aheap_1 \sep \aheap_2 
\\



\!\!\!\!\textsf{Perms}
& {\grammarPerm}
& {::=} & \fracperm \mid 1 \\
\!\!\!\!\textsf{Pure~formula}
& \pure & {::=} & \arith \mid \pure_1
{\wedge} \pure_2 \mid \pure_1 {\vee} \pure_2 \mid {\neg} \pure
\mid {\exists} \code{v} {\cdot} \pure \mid {\forall} \code{v} {\cdot} \pure\\

\!\!\!\!\textsf{Arith.~formula}
& \arith & {::=} & \aterm_1 {=} \aterm_2 \mid \aterm_1 {\neq} \aterm_2 \mid 
			\aterm_1 {<} \aterm_2 \mid \aterm_1 {\leq} \aterm_2  \\

\!\!\!\!\textsf{Arith.~term} 
& \aterm & {::=} & \code{k} \mid \code{v} \mid \code{k} \times \aterm \mid \aterm_1 + \aterm_2 \mid -\aterm \\

\end{array}
\]
\end{small}
\savespace
\[
\begin{array}{lll}
\!\!\! k {\in} \myit{integer~constants} &
\code{v} {\in} \myit{variables} , \vlist{v} \veq \code{v}_1,..,\code{v}_n
& 
~~\code{C} {\in} \myit{data~names}  
\\
\!\!\! \rsrvar {\in} \myit{resource~variables~} & 
\btt{R} {\in} \myit{resource~pred.~names}  & \sm{\fracperm}\in (0,1]

\end{array}
\]

\savesmallspace
\end{frameit}
\savespace
\caption{Core Specification Language}\label{fig:speclang}
\end{minipage}
 \savespace
\savespace
\savesmallspace
\end{center}
\end{figure}



\wnnay{to add defn for flow-aware predicates}

Fig.~\ref{fig:speclang} shows our specification language
for concurrent programs 
supporting \hidefa{``flow-aware''} {\CDL}'s abstract resource predicates. 
A classical separation logic formula $\constr$ is in disjunctive
normal form. Each disjunct in $\constr$ consists of
 formulae $\bheap$, $\heap$ and  $\pure$.
A pure formula $\pure$ includes standard equality/inequality and
Presburger arithmetic. 
$\pure$ could also be extended
to include other constraints such as set constraints.
The non-resource 
formula $\bheap$ may comprise action permission assertions \sm{\actionperm},  (permission annotated) \code{WAIT} relations, or
global views on shared states \sm{\GLOBAL{\code{v}}{\code{C}(\vlist{v})}}. 
The heap formula $\heap$ can be formed by simple heaps \code{\aheap}, 
\hidefa{(flow-annotated)} resource variables \code{\hidefa{[\fann]\,}\rsrvar}, 
resource predicate instances \sm{\btt{R}\hide{[\vlist{w}]}(\code{x},\overbar{\fconstr},\vlist{v})}, or via
separation conjunction $\sep$ (\cite{Ishtiaq:POPL01,Reynolds:LICS02}).
Note that a resource
variable \sm{\rsrvar} is a place holder for a 
formula \sm{\fconstr}, used in a resource predicate declaration, while
a resource predicate instance \sm{\btt{R}\hide{[\vlist{w}]}(\code{x},\overbar{\fconstr},\vlist{v})} encapsulates
\hidefa{either in-flow
or out-flow} resources  (\sm{\overbar{\fconstr}}) that are accessible via \code{x}.
For \hide{soundness and }precision reasons,
these \hidefa{flow-aware}resources are restricted to local entities that
can be tracked precisely, such as heap nodes or abstract
predicates, but must not include  global shared locations
or \code{\WAITK} relations.
A simple heap \code{\aheap} is formed by data nodes \sm{\permsto{\code{v}}{\code{C}(\vlist{v})}{}} (which can also appear in a thread-local view \sm{\LOCAL{\code{v}}{\code{C}(\vlist{v})}}).  

To allow us to focus on the essential issues, we  include limited support on permissions (namely on actions and \code{WAIT} relations) in our logic,
but this aspect can be relaxed to support more features of CSL \cite{Amighi:CoRR14}.
Furthermore, instances of \code{\WAITK} relations are only permitted
in pre/post specifications
(and not in predicate definitions). This is to allow every potential
race/deadlock errors to be detected by our lemmas, whenever it might occur.
Note also that, different from CAP \cite{Dinsdale-Young:ECOOP10} where each action is annotated with
a region, for simplicity we do not  explicitly mention regions in our action definition and the
shared region to be updated by an action can be recovered from the root pointers in the
action specification.
\scnay{To be consistent with the
following sentence, should $\fracperm$ be added to the points-to relation?}  
\hide{
The atomic heap formula
$\permsto{\term{v}}{ \thread{\constr} }{}$
(or thread node)
captures our idea of ``threads as resource'':
$v$ points to a thread carrying
certain resource $\constr$, which is available
after the thread is joined.
By representing threads as heap resource,
we allow them to be flexibly split and transferred
in a similar way as other types of resource such as heap nodes.
Note that thread nodes themselves are non-fractional,
since their resources can already be flexibly split.
We could allow fractional thread nodes; however,
it is always possible to rearrange heap nodes
into non-fractional forms (Section~\ref{sec:tas}).
Furthermore, no resource leakage from threads
is possible since we explicitly track when
each thread becomes dead. 

\khanh{the below semantics needs to be revised}
Our approach allows for expressive reasoning about threads
and their liveness.
For example, a formula  $\permsto{\term{t}}{ \thread{\constr} }{} \bigvee \dead{t} $
specifies the fact that the thread $t$ could be either alive or dead.
However, a formula
with $\permsto{\term{t}}{ \thread{\constr} }{} \wedge \dead{t}$
signifies an error state, because a thread cannot be both alive and dead.
Therefore, we enforce the requirement that concurrent threads
must maintain a program in thread-consistent (or \myit{t-consistent}) states.
}


\hide{
Thread nodes are non-fractional as they can be restructured.
}

\section{Automated Verification}
\label{sec:approach}
\wnnay{Khanh: Could you improve this section?}
Our verification system is built on top of entailment checking:
\savespace
\begin{equation*}
\entailCV{\D_{\myit{A}}}{\D_{\myit{C}}}{(\hvardef,\D_{\myit{R}})}
\savespace
\end{equation*}

\noindent
This entailment checks if antecedent $\D_{\myit{A}}$ is precise
enough to imply consequent $\D_{\myit{C}}$, and computes the residue $\D_{\myit{R}}$ 
for the next program state and resource bindings $\hvardef$.
$\hvardef$ is a set of pairs  $(\btt{V}, \constr_{\btt{V}})$
where $\btt{V}$ is a resource variable with its definition
$\constr_{\btt{V}}$.
\myit{E} is a set of existentially quantified variables from the consequent.
\hide{
Initially, the
entailment is invoked with \sm{\myit{E}=\emptyset},
in which case we abbreviate the entailment as
\entailH{\D_{\myit{A}}}{\D_{\myit{C}}}{(\hvardef,\D_{\myit{R}})}.
\myit{E} will be built up during the entailment.
}
This entailment procedure is used for
pre-condition and post-condition checking.
It is also used by assertion checking during 
automated verification.

In this section, we focus on the main entailment rules for 
manipulating resource predicates. The rest of the entailment
rules, such as those for manipulating normal data predicates 
or those for combining/normalizing resource 
predicates using lemmas introduced in Sec \ref{sec:flow},
are given in \techreport{Appendix \ref{app:entailment_rules}.
These other 
rules are adapted from prior works \cite{Nguyen:VMCAI07,Nguyen:CAV08} 
by additionally propagating the bindings $\hvardef$.
}{\cite{Chin:CDL2016}.}
Note that for simplicity, we omit fractional permissions 
from the predicates.

\hide{
The entailment rules
are presented in Fig.~\ref{fig:entail}.
\entrulen{[EX-L]} lifts existential variables out of
the entailment by replacing them with fresh variables.
\entrulen{[EX-R]} keeps track of existential variables
coming from the consequence in \myit{E}. After that,
the rest of entailment rules are applied to
quantifier-free formulas.
}

\hide{
{\noindent}{\bf Matching.}
Matching of two normal heap nodes (\entrulen{[MATCH]})
is accomplished by applying the substitution $\rho$ to
the remaining of the consequent.
The auxiliary function $\myit{freeEqn}$ is used to transfer
certain equations (for existential variables from
the consequent)
to the antecedent for subsequent
entailments. The following example illustrates the matching
of two cells:
\begin{equation*}
\frac{
  \entailH
      { \emp \wedge  v_1{=}v}
      { \emp }
      {(\emptyset, v_1{=}v)}
}
{
  \entailH
      { \permsto{\code{x}}{\code{cell}(v)}{} }
      { \permsto{\code{x}}{\code{cell}(v_1)}{} }
      {(\emptyset, v_1{=}v)}
}
\end{equation*}
}

{\noindent}{\bf Resource Predicate Matching.}
Matching of two resource predicates (\entrulen{[RP-MATCH]})
is the key of our approach, i.e. it allows resource predicates
to be split and identifies necessary resource bindings
for later entailments. For simplicity, we illustrate the rule
with at most one resource per predicate and we can handle
predicates with multiple resources by splitting them.

\begin{small}
\begin{equation*}
\frac{
\begin{array}{c}
\entrulen{[RP-MATCH]}\\
\rho = [\vlist{v}_1 / \vlist{v}_2]

\quad

(\rsrvar, \hide{\fann} \constr_{3}, \D, \code{b}) {=} 
\myit{addVar}(\rpreds{\btt{R}}{\code{x}, \hide{\fann} \rho(\constr_{2}),\vlist{v}_1})
\\

\hidefa{
\fann = \code{if \btt{R} is LatchIn then} ~ \OM ~ \code{else} ~ \OP
\quad}

\entailAnyVHO
    {\constr_{1}}
    {\constr_{3}}
    {(\hvardef, \rho_1)}
    {E \cup \{\rsrvar\}}
%



\\
\D_1 {=} \myit{subst}(\hvardef, \heap_1 {\wedge} \pure_1)

\quad

\D_2 {=} \myit{subst}(\hvardef, \D)

\quad 

\D_3 {=} \myit{subst}(\hvardef, \rho(\heap_2 {\wedge} \pure_2))

\\

\entailC{E \bagsubtract \{\vlist{v}_2\}}
	{\D_1 \sep \D_2 \wedge \myit{freeEqn}(\rho \cup \rho_1,E)}
	{\D_3}
  {(\hvardef_1,\D_4)}
  
\\

\hvardef_2 = \code{if b then} ~ \hvardef_1 ~ \code{else} ~ \hvardef \cup \hvardef_1

\end{array}
}
{
  \entailCV
      {\rpreds{\btt{R}}{\code{x}, \hide{\fann} \constr_{1}, \vlist{v}_1} \sep \heap_1 {\wedge} \pure_1}
      {\rpreds{\btt{R}}{\code{x}, \hide{\fann} \constr_{2},\vlist{v}_2} \sep \heap_2 {\wedge} \pure_2 }
      {(\hvardef_2,\D_4)}
}
\end{equation*}

\begin{equation*}
\begin{array}{lll}
\myit{addVar}(\rpreds{\btt{R}}{\code{x}, \hide{\fann} \rsrvar,\vlist{v}}) &\defs&
(\rsrvar, \hide{\fann} \rsrvar, \emp, \false)\\

\myit{addVar}(\rpreds{\btt{R}}{\code{x}, \hide{\fann} \heap {\wedge} \pure,\vlist{v}}) &\defs&
(\rsrvar, \hide{\fann} \heap {\sep} \rsrvar {\wedge} \pure, \rpreds{\btt{R}}{\code{x}, \hide{\fann} \rsrvar,\vlist{v}}, \true),
~\textit{fresh~\code{V}}
\end{array}
\end{equation*}

\vspace*{-4mm}

\begin{center}
\begin{multicols}{2}
\begin{equation*}
\hidefa{
\frac{
  \begin{array}{c}
    \entrulen{[RP-IN]}\\
    \entailCV
          {\constr_{2}}
          {\constr_{1}}
          {\hvardef}
  \end{array}
}{
  \entailInVHO
      {\constr_{1}}
      {\constr_{2}}
      {\hvardef}
      {\myit{E}}
}
\qquad
}
\frac{
\begin{array}{c}
    \entrulen{[RP-UNIFY]}\\
    \hide{
  \entailCV
          {\constr_{1}}
          {\constr_{2}}
          {\hvardef}
}
 \hvardef, \rho = {\it unify}(\constr_{1},\constr_{2})
  \end{array}
}{
  \entailAnyVHO
      {\constr_{1}}
      {\constr_{2}}
      {(\hvardef, \rho)}
      {\myit{E}}
}
\end{equation*}

\begin{equation*}
\hidefa{
\frac{
  \begin{array}{c}
    \entrulen{[RP-L-INST]}\\
    \rsrvar \in \myit{E}
  \end{array}
}{
\hide{
  \entailCV
            {\rsrvar}
            {\constr}
            {\{(\rsrvar, \constr)\}}
}
  \entailAnyVHO
            {\rsrvar}
            {\constr}
            {\{(\rsrvar, \constr)\}}
      {\myit{E}}
}
\qquad
}
\frac{
  \begin{array}{c}
    \entrulen{[RP-INST]}\\
    \rsrvar \in \myit{E}
  \end{array}
}{
\hide{  \entailCV
            {\constr}
            {\rsrvar}
            {\{(\rsrvar, \constr)\}}
}
  \entailAnyVHO
            {\constr}
            {\rsrvar}
            {(\{(\rsrvar, \constr)\}, \emptyset)}
      {\myit{E}}
}
\end{equation*}
\end{multicols}
\end{center}
\end{small}

In the above \entrulen{[RP-MATCH]} rule, we first apply the substitution $\rho$ to 
unify the components of the corresponding resource predicates 
in two sides of the entailment. 
We then resort to an \myit{addVar} method which either uses
an existing $\rsrvar$ (from the substituted RHS resource argument $\hide{\fann} \rho(\constr_{2})$) 
for binding, or adds
a fresh variable to the RHS to facilitate the 
implicit splitting of resource predicates (based on \code{\splitrulen{1}}
 and \code{\splitrulen{2}} rules).
After that, we invoke a special {\em resource} entailment
$
\entailAnyVHO
    {\constr_{1}}
    {\constr_{3}}
    {(\hvardef,\rho_1)}
    {E\cup \{\rsrvar\}}
$
for resource predicate $\btt{R}$, where
 the resource variable
$\rsrvar$ is added as an existential
variable to discover the resource bindings $\hvardef$ which maps 
resource variables to resource predicates (if any). Note that if $\rsrvar$
is a fresh  variable, such a binding about $\rsrvar$ will not
be kept in the final result.
We  also use two auxiliary functions: 
(i) \myit{subst} for substituting a discovered resource variable 
by its corresponding definition and 
(ii) $\myit{freeEqn}$ for transferring
certain equations (for existential variables from
the consequent) to the antecedent for subsequent
entailments.

\hide{
Rule \entrulen{[RP-IN]} 
ensures
that the inflow resource (i.e. \code{LatchIn}) is contravariant,
while \entrulen{[RP-OUT]} ensures that
the outflow resource (i.e. \code{LatchOut}) or normal resources are covariant. This
directly follows
from resource flow principles:
A resource \codeE{P} flows into another resource \codeE{Q}
if \codeE{P} is
more precise than (or entails) \codeE{Q}.
This use of flow-annotation to determine the
order of entailment is 
{\em critical} for ensuring accurate modeling of
the resource flows and exchanges.}
Our resource entailment discovers
resource bindings $\hvardef$ based on classic reasoning (without any frame residue)
via the rules \entrulen{[RP-UNIFY]} and \entrulen{[RP-INST]}.
Given two resource formulae $\constr_{1}$ and $\constr_{2}$, 
we first check if they are exact heaps by the rule \entrulen{[RP-UNIFY]} 
with unification where $\rho$ is the bindings of first-order variables. 
If there is any resource variable in the consequent formula,
we instantiate it by the rule \entrulen{[RP-INST]} after all common heaps 
in both sides are unified.
An example of the combination of matching (\entrulen{[RP-MATCH]}), followed
resource binding (\entrulen{[RP-INST]}) is: 
%
\[
\begin{array}{l}
\SPEC{ \LatchIn{c}{\,\permsto{\code{x}}{\code{cell}(v_1)}{}}} \SPEC{~\vdash~}  \SPEC{\LatchIn{c}{V}}
~\SPEC{\rewrite}~
\SPEC{ (\{(\code{V}, \permsto{\code{x}}{\code{cell}(v_1)}{})\},~~\code{\emp})}
\end{array}
\]
 
In many cases, resource predicates
are not matched but are rather split. In these cases,
a resource predicate with remaining resources 
(i.e. added by \myit{addVar}) is returned.
The predicate will be then added into the antecedent
in the rule \entrulen{[RP-MATCH]}. For example:
%
%
\[
\begin{array}{l}
\SPEC{\LatchOut{c}{\,\permsto{\code{x}}{\code{cell}(v_1)}{}{\sep}\permsto{\code{y}}{\code{cell}(v_2)}{}} }
\\
\SPEC{~\vdash~~}  \SPEC{\LatchOut{c}{\,\permsto{\code{x}}{\code{cell}(v_3)}{}} } 
\SPEC{~\rewrite~ (\sm{\emptyset}, \LatchOut{c}{\,\permsto{\code{y}}{\code{cell}(v_2)}{}} \wedge v_1{=}v_3) }
\end{array}
\]

\hide{
{\noindent}{\bf Splitting of Resource Predicates.}
As discussed above, splitting of resource-loaded predicates
(using lemmas such as \entrulen{[SPLIT-1]} and \entrulen{[SPLIT-2]})
is handled by the entailment rule \entrulen{[RP-MATCH]}.
This design decision allows for resource-directed splitting
(i.e. split only when there is a need to do so during matching)
and avoids the complication arising due to quantification of
resource variables.

Splitting of resource-less predicates (using lemmas such as
\entrulen{[SPLIT-3]} and \entrulen{[WAIT-3]}) is handled by the
rule \entrulen{[L-RP-SPLIT]}.
In order to apply a split lemma
$\aheap \wedge \pure \sm{\longrightarrow} \heap \in \lemmaset$ 
(where $\lemmaset$ is the set of lemmas presented in previous sections),
we find a substitution $\rho$
that matches $\aheap$ to
$\rpreds{\btt{R}}{\code{x}, \vlist{v}_1}$ and satisfies 
the guard constraint $\pure$.

{\noindent}{\bf Combining/Normalization of Resource Predicates.}
The combining/normalization
(using lemmas such as \entrulen{[NORM-1]} and \entrulen{[NORM-2]})
is handled by
the combine rule \entrulen{[L-RP-COMBINE]}.
In order to apply a combine lemma
$\aheap \sep \heap \wedge \pure \sm{\longrightarrow} \heap' \in \lemmaset$,
we find a substitution $\rho$ that matches $\aheap$ to
$\rpreds{\btt{R}}{\code{x}, \overbar{\fconstr}^1, \vlist{v}_1}$
and
satisfies $\heap \wedge \pure$.


{\noindent}{\bf Others.}
}


In order to provide the most precise program states,
we always perform 
normalization after each reasoning step.
\techreport{This normalization is performed with the help
of (\entrulen{[RP-COMBINE]}) in Appendix \ref{app:entailment_rules}.
}{} 


\section{Prototype Implementation}
\label{sec:implementation}

We demonstrate the feasibility of our approach in supporting
{\CDL} via \hidefa{flow-aware resource} 
concurrent abstract predicates
by implementing it
on top of \hip~which originally works
for only sequential programs.
\hide{
a recent 
verifier for concurrent programs
that is able to verify functional correctness 
(of heap-based programs) and deadlock-freedom.
{\parahip} models threads and coarse-grain locks,
but do not support flow-aware concurrent
predicates, or the more sophisticated
concurrency primitives, such as {\CDL}.
}
In contrast, besides verifying functional correctness,
data-race freedom, 
deadlock freedom, 
our new implementation (called {\toolname})
is now capable of also verifying 
{\CDL} and
beyond through its support for
concurrent abstract predicates,
shared local abstraction and wait-for relation.
Our {\toolname} prototype implementation and a set of
verified programs (involving {\CDL} and other
concurrency mechanisms)
are available for both online use
and download.\footnote{The URL is \blind{not given here due to the
requirement of double-blinded review}{at \toolurl}.} 

%
\hide{
The expressiveness of ``flow-aware resource predicate'' is beyond that
of other verification systems for fork/join programs.
However, as there is a lack of commonly accepted benchmarks in the literature,
we cannot easily compare {\toolname} with other systems.
In order to give readers an idea of the applicability of our approach,
we did an experimental comparison between {\sc ParaHIP}
and {\toolname}.
Experimental programs consist of 16 deadlock/deadlock-freedom
programs from {\sc ParaHIP}'s benchmark and other intricate programs
inspired by the literature.~\footnote{{\toolname} and
all experimental programs are available
\hide{for both  online use and download }at \toolurl.}
Besides the theoretical contributions,
the empirical questions we investigate are
(1) whether the
``flow-aware resource predicate'' approach
is capable of verifying more challenging programs,
and (2) how {\toolname} performs,
compared with {\sc ParaHIP}.
All experiments were conducted on a machine with
Ubuntu 12.04, 2.67GHz Intel$^{\circledR}$ Xeon$^{\circledR}$
X5650 processor, and 24GB memory.

\input{experiment}

Experimental results
are presented in Table~\ref{tab:result}.
{\toolname} is able to verify all programs
of {\sc ParaHIP}'s benchmark and other programs found
in the literature.
For these programs, {\sc ParaHIP} and {\toolname}
showed comparable verification times.
However, {\toolname} is more expressive than {\sc ParaHIP}
since it is 
capable of verifying more complex programs
that manipulate the multi-join pattern
and/or require expressive treatment of other
concurrency resources
such as \CDL~ and hierarchical locks

} 

\hide{
[Desbribe ParaHIP]
- Support separation logic
- MOdel threads as separate and-conjunctions, e.g.
hence, threads are transferred as a whole and can only joined once.
- Delayed lockset checking

Tips for implementation:
- Before discharge, 
 + Step1: Normalize first (syntatically combine all heap nodes of a single threads)
 + Step2: Infer separation (two separate nodes after separation are considered
separated)

For space reason, we show some examples. All examples show similar trend.
}






\vspace{-2mm}
\section{Related Work and Conclusion}
\label{sec:related}

Traditional works on concurrency verification
such as
Owicki-Gries \cite{Owicki:CACM76} and
Rely/Guarantee reasoning \cite{Jones:IFIP83}
are focused on using controlled interference to formally
reason between each process against those that execute
in the background.
Subsequent work on concurrency reasoning \cite{OHearn:CONCUR04} are based
mostly on race-freedom for concurrent processes by 
allowing processes to share read accesses on some 
locations, whilst having exclusive write accesses on others.
These works were conducted in the presence of
thread fork operations \cite{Hobor:thesis2011,Feng:ICFP2005}, 
and threads and locks \cite{Gotsman:APLAS07,Jacobs:POPL11,Leino:ESOP2010},
and use reasoning machineries, such as
RGSep \cite{Vafeiadis:CONCUR07},
LRG \cite{Feng:POPL09}, CAP \cite{Dinsdale-Young:ECOOP10},
and Views \cite{Dinsdale-Young:POPL2013},
but have not considered more advanced
concurrency synchronization, such as {\CDL}.
The closest to our work is a recent work on barrier
synchronization by Hobor and Gherghina 
\cite{Hobor:LMCS2012} which requires extra
pre/post specifications for each thread 
that are involved in each barrier synchronization
to allow resources to be 
exchanged.
This approach requires global inter-thread
reasoning to ensure that resources are preserved during each
barrier synchronization. In contrast, our use of
flow-aware concurrent predicates (for \CDL) rely on only modular
reasoning and would need a single set of
higher-order specifications for its
concurrency primitives. Moreover, we also ensure
race-freedom\hidefa{,
resource-preservation}
and deadlock-freedom.

\hide{
Even recent approaches such as CSL \cite{OHearn:CONCUR04},
RGSep \cite{Vafeiadis:CONCUR07},
LRG \cite{Feng:POPL09}, CAP \cite{Dinsdale-Young:ECOOP10},
and Views \cite{Dinsdale-Young:POPL2013},
omit fork/join concurrency from their languages.
\hide{Our ``flow-aware'' resource predicate, as a variant of CAP,
extends beyond these approaches to handle not only the multi-join pattern of first
class threads but also other popular concurrency mechanisms such as
\code{countDownLatch}.
}
Our ``flow-aware'' resource predicate is complementary to these approaches, and
our proposed approach not only  can handle first class threads well, but  is also capable of reasoning with
popular concurrency mechanisms such as
\code{countDownLatch}.

There also exist approaches that can handle fork/join
operations. Both Hobor~\cite{Hobor:thesis2011}
and Feng and Shao \cite{Feng:ICFP2005}
support fork and omit join with the claim that
thread join can be implemented using synchronization. 
However, without join, the former allows threads
to leak resource upon termination
while the latter requires global specifications of
inter-thread interference.
Approaches that can handle both fork and join can
be grouped as modeling
threads as tokens \cite{Gotsman:APLAS07,Jacobs:POPL11,Leino:ESOP2010}
and modeling threads as \lit{and}-conjuncts
\cite{Le:ATVA13,Le:ICFEM2012}.
Though syntactically different, they are semantically similar
in that the tokens and the \lit{and}-conjuncts are used
to represent the post-states of forked threads.
However, they offer limited support for first-class threads:
tokens and \lit{and}-conjuncts are not allowed
to be split and shared among concurrent threads.
As such, they are not expressive enough
to verify programs with more intricate fork/join
behaviors such as the multi-join pattern
where threads are shared
and joined in multiple threads. Note that the multi-join pattern illustrated in Fig~\ref{fig:mapreduce} earlier also appears in the
context of asynchronous tasks.
}

\hide{
Inspired by the key notation of resource in separation
logic \cite{Bornat:ENTCS06,OHearn:CONCUR04}, we propose
to model threads as resource, thus allow ownerships
of threads to be flexibly split and distributed
among multiple joiners.
This enables verification of the
multi-join pattern.\hide{
In addition, unlike ours, none of related works that
we are aware of support explicit reasoning about
thread liveness.}
To the best of our knowledge, only 
Haack and Hurlin \cite{Haack:AMAST08}
can reason about some multi-join scenarios.
In their approach, a thread token can be
associated with a fraction and this allows
multiple joiners to join with the same joinee
in order to read-share the joinee's resource.
However, write-share is not permitted
unlike our use of flow-aware predicates.
}


Our 
resource predicates are 
based on 
Concurrent Abstract Predicates (CAP) 
\cite{Dinsdale-Young:ECOOP10,Dodds:POPL11,Svendsen:ESOP13,Svendsen:ESOP14}.  
\hide{Dinsdale-Young et al. \cite{Dinsdale-Young:ECOOP10} 
first propose CAP to help give abstract specifications for concurrent data structures so that
client programs can refer to such CAPs without the need to know
the actual implementations of these concurrent data structures.} 
The basic idea behind 
CAP \cite{Dinsdale-Young:ECOOP10}  was to provide
an abstraction of possible interferences from concurrently running threads, by 
partitioning the state into regions with protocols governing how the state in each region is 
allowed to evolve. 
Dodds et al. \cite{Dodds:POPL11} introduced a higher-order variant of CAP to give a generic
specification for a library for deterministic parallelism, making explicit use of nested region assertions
and higher-order protocols. 
Svendsen et al. \cite{Svendsen:ESOP13} presented  a  logic called Higher Order Concurrent
Abstract Predicates (HOCAP), allowing clients to refine the generic specifications of concurrent data structures.  
\hide{HOCAP was 
developed based on Jacobs and Piessens' idea of parameterizing specifications of concurrent methods  with ghost code,
to be executed in synchrnonization points  \cite{Jacobs:POPL11}. But instead of resorting to lock invariants from clients
as in \cite{Jacobs:POPL11}, }%
HOCAP uses higher order (predicative) protocols to allow clients to  transfer
ownership of additional resources to shared data structures.  

More recently, there have been a number of modern concurrency logics such as iCAP  \cite{Svendsen:ESOP14},  Iris \cite{popl15:iris}, FCSL\cite{DBLP:conf/esop/NanevskiLSD14}\cite{DBLP:conf/pldi/SergeyNB15} and CoLoSL \cite{DBLP:conf/esop/RaadVG15}. 
As an improved version of 
\cite{Svendsen:ESOP13}, iCAP \cite{Svendsen:ESOP14}   allows the use of
impredicative protocols parameterised on arbitrary predicates and supports modular reasoning about layered and recursive abstractions.  
Iris \cite{popl15:iris}  combines partial commutative monoids (PCMs) and invariants.  Both Iris and iCAP leverage the idea of view shifts, originated by Jacobs and Piessens  \cite{Jacobs:POPL11}. FCSL \cite{DBLP:conf/esop/NanevskiLSD14,DBLP:conf/pldi/SergeyNB15},  incorporates a uniform concurrency model, based on state-transition systems and PCMs,  so as to build proofs about concurrent libraries in a thread-local, compositional way. CoLoSL \cite{DBLP:conf/esop/RaadVG15} allows each thread to be verified with respect to its partial subjective view of the global shared state, and uses overlapping conjunction \cite{DBLP:conf/popl/HoborV13} to reconcile the permissions and capabilities, residing in the shared state between concurrent threads.
%
%
%
Compared with these general frameworks, our 
abstract predicates are  resource-specific as they explicitly track resources that flow into and out of their abstractions and allow resources to be flexibly
split and transferred across procedure and thread boundaries. 
\hide{
Our proposed specification and verification mechanism is rather general as it 
not only supports
first-class threads, but is also capable of handling other popular concurrency patterns such as  \code{countDownLatch},
locks and copyless message passing \cite{Leino:ESOP2010,Villard:APLAS09}.
}
To deal with the shared counter mechanism, we have
now proposed a thread-local abstraction which makes
it much more precise and yet simpler to ensure stability of 
our 
concurrent abstract predicates.
We have also used {\contra}  lemma support
to ensure deadlock freedom and race-freedom,
and have used  resource predicates 
to help ensure resource preservation -- desirable properties that
were not properly addressed by prior work on CAP.
In another research direction, deadlock avoidance by verification
were recently investigated for locks
and channels \cite{Leino:ESOP2010,Le:ATVA13}.
In comparison, we have provided a new approach
based on  {\contra} lemma and wait-for set
to ensure deadlock freedom.




\hide{
As a part of our future work, we are also working
towards using flow-aware resource predicates to
support automated verification of communication protocol,
such as those addressed by session
types \cite{Honda:POPL08} and contracts \cite{Villard:APLAS09}. The \sm{\OM} annotation
corresponds to send \sm{!} annotation,
while \sm{\OP} corresponds to receive \sm{?}
annotation for both session types and contracts.
Unlike these past works, we are placing
importance on using contravariant and covariant subsumption
principles to support first-class communications.
Compared to types, we will be able to leverage
on full expressivity of separation logic.
Compared to global contracts, our use of flow-aware
concurrent predicates could
support better localised reasoning.
}


\label{sec:concl}


In conclusion, we have proposed a framework
to validate the correctness
of concurrent programs using {\CDL}.
We showed
how to ensure {\em race-freedom} 
and {\em deadlock-freedom}.
To the best of our 
knowledge, this is 
the first proposal on
formal verification for the correctness of 
concurrent programs using {\CDL}.
We have made use of 
{\em concurrent abstract predicates}
to precisely track resources that are exchanged via
\CDL.
Our proposal allows tracked resources to
be re-distributed,
in support of 
sharing and 
synchronization amongst a group of
concurrent threads. 
We have also proposed 
{\contra} lemmas
to assist with deadlock and race detection.
We have followed the approach of \cite{Dinsdale-Young:ECOOP10}
for verifying the correctness of an
implementation for {\CDL}, but requires two new
concepts (i) borrowed heap
via \code{P{\septract}\emp} whereby \codeS{$\forall \code{Q} {\cdot} (\code{Q}{\septract}\emp){\sep}\code{Q} = \emp$}, and (ii) 
thread-local abstraction
for precise tracking of shared counters.
\hide{
We have also formalized several desirable properties
of flow-aware concurrency reasoning. Firstly, we are 
able to guarantee race freedom, since each resource is
being tracked precisely. Secondly, we can also guarantee
deadlock-freedom for single-resource scenarios with the help
of only synchronization error lemmas. If multiple resources
are being used, 
we can add a waits-for relation to determine
a partial ordering\hide{
(\cite{Leino:ESOP2010} requires pre-declared total ordering)}
on the resources. Absence of a cycle
in the ordering guarantees deadlock freedom.
Lastly, we show how to ensure resource-preservation for each
of our program units.
}
\hide{We have implemented our approach in a tool,
called {\toolname},
to verify partial correctness,
data-race freedom, deadlock-freedom, and resource-preservation
for concurrent programs.
}
Lastly, proof of soundness of our verification
framework is still being developed. It is 
based the Views framework \cite{Dinsdale-Young:POPL2013} for modular
concurrency reasoning, but have to be 
extended to cater to several new features
, namely
(i) inconsistency lemmas
(ii) borrowed heap
(iii) shared local abstractions,
and (iv) deadlock freedom guarantee.

\noindent
{\bf Acknowledgement:} We gratefully acknowledge
Duy-Khanh Le who highlighted this 
\CDL~problem to us and helped with our initial
formulation based on flow-aware predicates.


\bibliographystyle{abbrv} 
\bibliography{main-short}


\newpage
\appendix

\section{Additional Entailment Rules}
\label{app:entailment_rules}
\begin{small}
 \scnay{Chanh: are these rules up-to-date?}

Additional entailment rules are given in Fig~\ref{fig:entail}.
\begin{figure}[h]
\begin{center}
\begin{frameit}
\begin{small}
\savespace
\savespace
\savesmallspace
\begin{equation*}
\frac{
\begin{array}{c}
\entrulen{[EX-L]}\\
\myit{fresh}~\code{w}

\quad

\entailCV{[\code{w/v}]\D_1}
         {\D_2}
         {(\hvardef,\D)}
         
\end{array}
}
{
  \entailCV{{\exists} \code{v} {\cdot} \D_1}
           {\D_2}
           {(\hvardef,\D)}
}
\end{equation*}
\\
\begin{equation*}
\frac{
\begin{array}{c}
\entrulen{[EX-R]}\\
\myit{fresh}~\code{w}

\quad

\entailC{E \bagunion {\code{w}}}
        {\D_1}
        {[\code{w/v}]\D_2}
        {(\hvardef,\D_3)}

\quad

\D \defs {\exists} \code{w} {\cdot} \D_3

\\
\end{array}
}
{
  \entailCV
      {\D_1}
      {{\exists} \code{v} {\cdot} \D_2}
      {(\hvardef,\D)}
}
\end{equation*}

\begin{equation*}
\frac{
\begin{array}{c}
\entrulen{[MATCH]}\\
\rho = [\vlist{v}_1 / \vlist{v}_2]
\\
\entailC{\myit{E}-\{\vlist{v}_2\}}
	{\heap_1 {\wedge} \pure_1 {\wedge} \myit{freeEqn}(\rho,\myit{E})}
	{\rho(\heap_2 {\wedge} \pure_2)}
        {(\hvardef,\D)}
\end{array}
}
{
  \entailCV
      {\permsto{\code{x}}{C(\vlist{v}_1)}{} \sep \heap_1 {\wedge} \pure_1}
      {\permsto{\code{x}}{C(\vlist{v}_2)}{} \sep \heap_2 {\wedge} \pure_2}
      {(\hvardef,\D)}
}
\end{equation*}

\hide{
\begin{minipage}{15pc}
\begin{equation*}
  \begin{array}{c}
    \entrulen{[RP-2]}\\
    \entailHO
        {\rsrvar {\wedge} \true}
        {\heap {\wedge} \pure}
        {( \{(\rsrvar, \heap {\wedge} \pure)\}, \emp )}
  \end{array}
\end{equation*}
\vspace*{-2ex}
\begin{equation*}
  \begin{array}{c}
    \entrulen{[RP-3]}\\
    \entailHO
        {\heap {\wedge} \pure}
        {\rsrvar {\wedge} \true}
        {( \{(\rsrvar, \heap {\wedge} \pure)\}, \emp )}
  \end{array}
\end{equation*}
\end{minipage}
\begin{minipage}{14pc}
\begin{center}
\begin{equation*}
\frac{
  \begin{array}{c}
    \entrulen{[RP-4]}\\
    \entailC
      {\emptyset}
      {\heap_1 {\wedge} \pure_1}
      {\heap_2 {\wedge} \pure_2}
      {(\hvardef,\D)}
  \end{array}
}{
  \entailHO
      {\heap_1 {\wedge} \pure_1}
      {\heap_2 {\wedge} \pure_2}
      {( \hvardef, \D )}
}
\end{equation*}
\end{center}
\end{minipage}
}

\hide{
\begin{equation*}
\frac{
\begin{array}{c}
\entrulen{[L-SPLIT]}\\
\aheap \wedge \pure \sm{\longrightarrow} \heap \in \lemmaset
\qquad
\rho = match(\aheap, \permsto{\code{x}}{C(\vlist{v}_1)}{})
\\
\entailS{\permsto{\code{x}}{C(\vlist{v}_1)}{} \sep \heap_1 {\wedge} \pure_1 }
        { \rho(\pure)}
\qquad
\entailCV{\rho(\heap) \sep \heap_1 {\wedge} \pure_1}
         {\heap_2 {\wedge} \pure_2}
         {(\hvardef,\D)}

\end{array}
}
{
  \entailCV
      {\permsto{\code{x}}{C(\vlist{v}_1)}{} \sep \heap_1 {\wedge} \pure_1}
      {\heap_2 {\wedge} \pure_2}
      {(\hvardef,\D)}
}
\end{equation*}

\begin{equation*}
\frac{
\begin{array}{c}
\entrulen{[L-COMBINE]}\\
\aheap \sep \heap \wedge \pure \sm{\longrightarrow} \heap' \in \lemmaset
\qquad
\rho = match(\aheap, \permsto{\code{x}}{C(\vlist{v}_1)}{})
\\
\entailCV{\permsto{\code{x}}{C(\vlist{v}_1)}{} \sep \heap_1 {\wedge} \pure_1 }
        { \rho(\heap \wedge \pure)}
        {(\hvardef_1,\D_1)}
\qquad
\entailCV{\rho(\heap') \sep \D_1}
         {\heap_2 {\wedge} \pure_2}
         {(\hvardef_2,\D)}

\end{array}
}
{
  \entailCV
      {\permsto{\code{x}}{C(\vlist{v}_1)}{} \sep \heap_1 {\wedge} \pure_1}
      {\heap_2 {\wedge} \pure_2}
      {(\hvardef_1 \bagunion \hvardef_2,\D)}
}
\end{equation*}
} 

\hide{ 
\begin{equation*}
\frac{
\begin{array}{c}
\entrulen{[L-RP-SPLIT]}\\
\aheap \wedge \pure \sm{\longrightarrow} \heap \in \lemmaset
\qquad
\rho = match(\aheap, \rpreds{\btt{R}}{\code{x}, \vlist{v}_1})
\\
\entailS{\rpreds{\btt{R}}{\code{x}, \vlist{v}_1} \sep \heap_1 {\wedge} \pure_1 }
        { \rho(\pure)}
\qquad
\entailCV{\rho(\heap) \sep \heap_1 {\wedge} \pure_1}
         {\heap_2 {\wedge} \pure_2}
         {(\hvardef,\D)}

\end{array}
}
{
  \entailCV
      {\rpreds{\btt{R}}{\code{x}, \vlist{v}_1} \sep \heap_1 {\wedge} \pure_1}
      {\heap_2 {\wedge} \pure_2}
      {(\hvardef,\D)}
}
\end{equation*}
} 

\begin{equation*}
\frac{
\begin{array}{c}
\entrulen{[RP-COMBINE]}\\
\aheap \sep \heap \wedge \pure \sm{\longrightarrow} \heap' \in \lemmaset
\qquad
\rho = match(\aheap, \rpreds{\btt{R}}{\code{x}, \overbar{\fconstr}^1, \vlist{v}_1})
\\
\entailCV{\rpreds{\btt{R}}{\code{x}, \overbar{\fconstr}^1, \vlist{v}_1} \sep \heap_1 {\wedge} \pure_1 }
        { \rho(\heap \wedge \pure)}
        {(\hvardef_1,\D_1)}

\\

\entailCV{\rho(\heap') \sep \D_1}
         {\heap_2 {\wedge} \pure_2}
         {(\hvardef_2,\D)}

\end{array}
}
{
  \entailCV
      {\rpreds{\btt{R}}{\code{x}, \overbar{\fconstr}^1, \vlist{v}_1} \sep \heap_1 {\wedge} \pure_1}
      {\heap_2 {\wedge} \pure_2}
      {(\hvardef_1 \bagunion \hvardef_2,\D)}
}
\end{equation*}

\begin{equation*}
\begin{array}{c}
\myit{freeEqn}([\code{u}_i/\code{v}_i]_{i=1}^n,\myit{V}) \defs \\
\code{let} ~ \pure_i = (\code{if}~ \code{v}_i {\in} \myit{V} ~ \code{then} ~ \true ~ \code{else} ~ \code{v}_i = \code{u}_i) ~ \code{in} ~\bigwedge_{i=1}^n \pure_i 
\end{array}
\end{equation*}
\hide{ 
\begin{equation*}
\begin{array}{c}
\!\!\myit{isEmp}(\emp) \defs \true \quad
\myit{isEmp}(\heap {\wedge} \pure) \defs \myit{isEmp}(\heap) \quad
\myit{isEmp}(\heap_1 {{\sep}} \heap_2) \defs \myit{isEmp}(\heap_1){\wedge} \myit{isEmp}(\heap_2)\\
\myit{isEmp}(\D_1 \vee \D_2) \defs \myit{isEmp}(\D_1) {\wedge} \myit{isEmp}(\D_2) \qquad
\myit{isEmp}(\aheap) \defs \false ~~\myit{for}~\myit{atomic heap}~ \aheap
\end{array}
\end{equation*}
} 

\begin{equation*}
\begin{array}{c}
\myit{match} (\rpreds{\btt{R}}{\code{x}_1, \overbar{\fconstr}^1 , \vlist{v}_1}, \rpreds{\btt{R}}{\code{x}_2, \overbar{\fconstr}^2, \vlist{v}_2})
\defs
[\code{x}_1/\code{x}_2, \vlist{v}_1/\vlist{v}_2] 
\end{array}
\end{equation*}

\end{small}
\end{frameit}
\savesmallspace
\caption{Additional Entailment Rules}\label{fig:entail}
\end{center}
\end{figure}

 \end{small}

\section{A Thread-Local Abstraction for \CDL}

\section{A Verified {\CDL} Implementation}
With the interpretations for the
flow-aware predicates, we can now prove
soundness of our normalization,
splitting and {\contra}
lemmas.


\hide{
\scsay{The proof for \splitrulen{2} need patching}

\begin{proof}[\splitrulen{2}]
\begin{small}
\[
\begin{array}{l}
\LatchIn{i}{(P{\sep}Q)}
\\
\Leftrightarrow {[\DEC]}_{\fracpermc} {\sep} {\OM} (\code{P} {\sep} \code{Q}) {\sep}\GLOBAL{i}{n} {\wedge}\code{n>0}
\\
\Rightarrow {[\DEC]}_{\fracperm_1} {\sep}   {[\DEC]}_{\fracperm_2} {\sep} {\OM} \code{P} {\sep} {\OM} \code{Q} {\sep}\GLOBAL{i}{n}{\sep}\GLOBAL{i}{n} {\wedge}\code{n>0}{\wedge}\fracpermc=\fracperm_1{+}\fracperm_2
\\
\Rightarrow \LatchIn{i}{P}  {\sep} \LatchIn{i}{Q}
\end{array}
\]
\end{small}
\end{proof}
}
\hide{
}
\hide{

\begin{verbatim}
Need to change:
  lc(i,-P) == DEC{f}*-P*[i->n]&n>0

Proofs for Splitting Lemmas
===========================
  // lc(i,-P*Q)
  // DEC{f}*-(P*Q)*[i->n]&n>0 & 0<f<=1
  // DEC{f1}*DEC{f2}*-P*-Q*[i->n]&&0<f1,f2<=1 & f=f1+f2 & n>0
  lc(i,-P*Q) --> lc(i,-P)*lc(i,-Q) 
  // DEC{f1}*-P*[i->n]*DEC{f2}*-Q*[i->n]&0<f1,f2<=1 & f=f1+f2 & n>0
  // lc(i,-P)*lc(i,-Q) 

  // lc(i,+P*Q)
  // +(P*Q)
  lc(i,+P*Q) --> lc(i,+P)*lc(i,+Q) 
  // P*Q)
  // lc(i,+P)*lc(i,+Q) 

  // CNT(i,n) & n1,n2>=0 & n=n1+n2 
  // {|i->n|} & n1,n2>=0 & n=n1+n2
  // {|i->n1|}*{|i->n2|} & n1,n2>=0 & n=n1+n2
  CNT(i,n) & n1,n2>=0 & n=n1+n2 --> CNT(i,n1)*CNT(i,n2)
  // {|i->n1|}*{|i->n2|} & n1,n2>=0 & n=n1+n2
  // CNT(i,n1)*CNT(i,n2) & n1,n2>=0 & n=n1+n2
  // CNT(i,n1)*CNT(i,n2)

Proofs for Normalization Lemmas
===============================
  // CNT(c,n) * CNT(c,-1) & n<=0 
  // {|c->n|}& n<=0 * {|c->-1|}
  // {|c->n|}& (n=-1|n=0) * [c->0]
  // ({|c->-1|} | {|c->0|})  * [c->0]
  // ([c->0] | [c->m]&m>=0)  * [c->0]
  CNT(c,n) * CNT(c,-1) & n<=0 ==> CNT(c,-1) // idempotent lemma 
  // ([c->0] | [c->0])
  // [c->0] 
  // CNT(c,-1) 


  // CNT(c,n1) * CNT(c,n2) & n=n1+n2 & n1,n2>=0 & n>=0  
  // {|c->n1|}*{|c->n2|} & n=n1+n2 & n1,n2>=0 & n>=0 
  // {|c->n1+n2|} & n=n1+n2 & n1,n2>=0 & n>=0 
  CNT(c,n1) * CNT(c,n2) & n=n1+n2 & n1,n2>=0 & n>=0  ==> CNT(c,n)
  // {|c->n|} & n=n1+n2 & n1,n2>=0 & n>=0 
  // CNT(c,n) & n=n1+n2 & n1,n2>=0 & n>=0 
  // CNT(c,n)

Proofs for Contradiction Lemmas
===========================================
  // lc(c,-P) * CNT(c,-1) & a>0 
  // DEC{f}* {|c->n>0|} * {|c->-1|} & a>0 
  // DEC{f}* [c->m] * [c->0] & m>0 
  lc(c,-P) * CNT(c,-1)   --> RACE-ERROR // race problem
  // [c->m /\ c->0] & m>0 
  // false
  // RACE-ERROR

  // CNT(c,a) * CNT(c,-1) & a>0 
  // {|c->a|} * {|c->-1|} & a>0 
  // [c->m] * [c->0] & m>=a>0 
  CNT(c,a) * CNT(c,-1) & a>0  --> DEADLOCK  // wait without any grant
  // [c->m /\ c->0] & m>=a>0 
  // false
  // DEADLOCK  
\end{verbatim}
}
\hide{
\hide{
To support multi-latch deadlocks, we have
also provided a \cc{WAIT(S)} set which is used to
capture waiting of the form \cc{c2{\ra}c1}, 
where \cc{c2} must complete before \cc{c1}.
In case we detect a cycle in such
wait-for arcs in \cc{S}, it would signal
a possible deadlock sceanario where a set of threads
are involved in some cyclic waiting.

However, if the waits-for set is cycle-free, we are
guaranteed to have some concurrency
schedule that would complete all the waiting
threads, based on the ordering of this waits-for
graph. This lemma cannot be proven as
a {\contra} lemma since it is not based
on some heap states. Instead, we have relied
on the expected ordering of execution to ensure
deadlock-freedom.
}
}

We must also verify the correctness of
{\CDL} implementation. 
A more realistic implementation will make use of
locks, wait and notifyAll operations to implement
blocking of {\await} commands until the shared
counter reaches 0. 
For ease of presentation,
we have used a simple
version to illustrate how thread-local abstraction
was used, as shown below.

$
\begin{array}{l}
\code{{\CDL}\, {\createlatch}(int\, n) \{}
\code{return new \CDL(n); \}}
\\
\code{void\,countDown({\CDL} i) \{ }
\code{\criticalSec{if (i.val>0) i.val = i.val-1;\hide{else notifyAll();}}\ \}}
\\
\code{void await({\CDL} i) \{ while (i.val>0) skip; \}}
\end{array}
$

\hide{
We highlight the verification steps for
one pre/post specification of {\CD}. The rest are left 
in Appendix~\ref{app:verified-methods}.

\hide{
$
\begin{array}{l}
\code{{\CDL}\, {\createlatch}(int\, n) \code{with} \code{P} }
\\
\indent\quad \SPEC{\requires~\code{n\sm{\gt}0}} \\
\indent\quad \SPEC{\ensures~ \LatchIn{res}{P} {\sep}\LatchOut{res}{P}{\sep}\code{CNT(res,n)}} \\
\code{\{}
\\
\qquad \CSPEC{\code{n\sm{\gt}0}}
\\
\qquad
\code{{\CDL}\,i = new {\CDL}(n);}
\\
\qquad \CSPEC{i\sm{\mapsto}\CDL(n)\sm{\wedge}\code{n\sm{\gt}0}}
\\ 
\qquad \CSPEC{\sm{{[\DEC]}_{1} {\sep} \OP \code{P} {\sep} \OM \code{P} {\sep} \GLOBAL{i}{n}}\sm{\wedge}\code{n\sm{\gt}0}}
\\ 
\qquad \CSPEC{\sm{{[\DEC]}_{1} {\sep} \OP \code{P} {\sep} \OM \code{P} {\sep} \GLOBAL{i}{n} {\sep} \LOCAL{i}{n} {\wedge}\code{n\sm{\gt}0}} }
\\
\qquad \CSPEC{\sm{\LatchIn{i}{P} {\sep} \LatchOut{i}{P} {\sep} \CNT{i}{n}}}
\\
\qquad \code{return i; }
\\
\qquad \CSPEC{\sm{\LatchIn{\res}{P} {\sep} \LatchOut{\res}{P} {\sep} \CNT{\res}{n}}}
\\
\code{\}}
\end{array}
$
\\[1ex]
\indent
$
\begin{array}{l}
\code{{\CDL}\, {\createlatch}(int\, n) \code{with} \code{P} }
\\
\indent\quad \SPEC{\requires~\code{n\sm{=}0}} \\
\indent\quad \SPEC{\ensures~ \code{CNT(res,-1)}} \\
\code{\{}
\\
\qquad \CSPEC{\code{n\sm{=}0}}
\\
\qquad
\code{{\CDL}\,i = new {\CDL}(n);}
\\
\qquad \CSPEC{i\sm{\mapsto}\CDL(n)\sm{\wedge}\code{n\sm{=}0}}
\\ 
\qquad \CSPEC{\GLOBAL{i}{n}\sm{\wedge}\code{n\sm{=}0}}
\\
\qquad \CSPEC{\CNT{i}{-1}}
\\
\qquad\code{return i; }
\\
\qquad \CSPEC{\CNT{res}{-1}}
\\
\code{\}}
\end{array}
$
\\[1ex]
\indent
}

$
\begin{array}{l}
\code{void~countDown(\CDL~i)}\\
\indent\quad \SPEC{\requires~\LatchIn{i}{P}{\sep}\code{P}{\sep}\code{CNT(i,n)}\sm{\wedge}\code{n{>}0}}\\
\indent\quad \SPEC{\ensures~ \code{CNT(i,n{-}1)}};\\ 
\code{\{}
\\
\qquad \CSPEC{\LatchIn{i}{P}{\sep}\code{P}{\sep}\code{CNT(i,n)}\sm{\wedge}\code{n{>}0}}
\\
\qquad \CSPEC{\sm{(\code{P} {\septract}\emp){\sep}~{[\DEC]}_{\fracpermc} {\sep} \GLOBAL{i}{m} {\wedge}\code{m{>}0}} {\sep}\,\code{P}\,{\sep} \LOCAL{i}{n} \sm{\wedge}\code{n\sm{>}0}}
\\
\qquad \CSPEC{\sm{{[\DEC]}_{\fracpermc} {\sep}\GLOBAL{i}{m} {\wedge}\code{m{>}0}} {\sep} \LOCAL{i}{n} \sm{\wedge}\code{n\sm{>}0}}
\\
\qquad \code{\criticalSec{if (i.val>0) i.val=i.val-1;}}
\\
\qquad \CSPEC{\sm{\GLOBAL{i}{m{-}1} {\wedge}\code{m{>}0} } {\sep} \LOCAL{i}{n{-}1} \sm{\wedge}\code{n\sm{>}0}}
\\
\qquad \CSPEC{\LOCAL{i}{n{-}1} \sm{\wedge}\code{n\sm{>}0}}
\\
\qquad \CSPEC{\CNT{i}{n{-}1}}
\\
\code{\}}
\end{array}
$

\hide{
\\[1ex]
\indent
$
\begin{array}{l}
\code{void~countDown(\CDL~i)}\\
\indent\quad \SPEC{\requires~\code{CNT(i,-1)}}\\
\indent\quad \SPEC{\ensures~ \code{CNT(i,-1)}};\\
\code{\{}
\\
\qquad \CSPEC{\code{CNT(i,-1)}}
\\
\qquad \CSPEC{\GLOBAL{i}{0}}
\\
\qquad\code{\criticalSec{if (i.val>0) i.val=i.val-1;}}
\\
\qquad \CSPEC{\GLOBAL{i}{0}}
\\
\qquad \CSPEC{\code{CNT(i,-1)}}
\\
\code{\}}
\end{array}
$
\\[1ex]
\indent
$
\begin{array}{l}
\code{void~await(\CDL~i)}\\
\indent\quad \SPEC{\requires~\LatchOut{i}{P}{\sep}\code{CNT(i,0)}}\\
\indent\quad \SPEC{\ensures~ \code{P}{\sep}\code{CNT(i,-1)}};\\
\code{\{}
\\
\qquad \CSPEC{\LatchOut{i}{P}{\sep}\code{CNT(i,0)}}
\\
\qquad \CSPEC{\code{P}{\sep}\LOCAL{i}{0}}
\\
\qquad \CSPEC{\code{P}{\sep}\GLOBAL{i}{m}\sm{\wedge}\code{m\sm{\geq}0}}
\\
\qquad \code{while (i.val>0) skip; }
\\
\qquad \CSPEC{\code{P}{\sep}\GLOBAL{i}{0} }
\\
\qquad \CSPEC{\code{P}{\sep}\code{CNT(i,-1)}}
\\
\code{\}}
\end{array}
$

\indent
$
\begin{array}{l}
\code{void~await(\CDL~i)}\\
\indent\quad \SPEC{\requires~\code{CNT(i,-1)}}\\
\indent\quad \SPEC{\ensures~\code{CNT(i,-1)}};\\
\code{\{}
\\
\qquad \CSPEC{\code{CNT(i,-1)}}
\\
\qquad \CSPEC{\GLOBAL{i}{0}}
\\
\qquad \code{while (i.val>0) skip; }
\\
\qquad \CSPEC{\GLOBAL{i}{0}}
\\
\qquad \CSPEC{\code{CNT(i,-1)}}
\\
\code{\}}
\end{array}
$
}
}
\hide{
\begin{verbatim}
  CDL newCDL(n) with P
   requires n>=0  // exception thrown when n is negative
   ensures lc(res,+P) * lc(res,-P) * CNT(res,n)
  {
    // emp
    i=alloc(1)
    // i->_
    i.val=n;
    // i->n
    // DEC * P * -P * [i->n] & n>=0
    // DEC * P * -P * {|i->n|}
    // lc(i,+P)*lc(i,-P)*CNT(i,n)
    return i;
    // lc(res,+P)*lc(res,-P)*CNT(res,n)
  }

 void countDown(CDL i) 
   requires lc(i,-P)*P*CNT(i,n) & n>0
   ensures  CNT(i,n-1);
   requires CNT(i,-1) 
   ensures  CNT(i,-1);
 {
   // lc(i,-P)* P * CNT(i,n) & n>0
   // [i->c] & c>0 * -P * P * {|i->n|} & n>0 
   // [i->c] & c>0 * {|i->n|} & n>0
   // [i->c] & c>0 * {|i->n|} & n>0
    <if i.val>0 then i.val=i.val-1>
   // [i->c-1] & c>0 * {|i->n-1|} & n>0
   // {|i->n-1|} & n>0
   // CNT(i,n-1) & n>0
   // CNT(i,n-1) 
 }
   requires CNT(i,-1) 
   ensures  CNT(i,-1);
 {
   // CNT(i,-1) 
   // {|i->-1|} 
   // [i->0] 
    <if i.val>0 then i.val=i.val-1>
   // [i->0] 
   // {|i->-1|}
   // CNT(i,-1)
 }

  void await(i)
   requires lc(i,+P) * CNT(i,n) & n<=0
   ensures  P * CNT(i,-1);
   {
   // lc(i,+P) * CNT(i,n) & n<=0
   // P *{|i->n|} & n<=0
   // P *{|i->n|} & (n=-1  | n=0)
   // P * ([i->0] | [i->m] & m>=0 & n=0)
   while(i.val>0)
     skip; 
   // P * [i->0] 
   // P * {(|i->-1|} 
   // P*CNT(i,-1)
  }
\end{verbatim}
}

\section{Proof of Lemmas for a CDL Library}
\label{app:verified-lemmas}
With the interpretations for the
flow-aware predicates, we can now prove
soundness of our normalization,
splitting and {\contra} lemmas, as follows:

\scnay{The proofs for \splitrulen{1},\splitrulen{2} and \errrulen{1} require patching}

\begin{proof}[\splitrulen{1}]
\begin{small}
\[
\begin{array}{l}
\LatchOut{i}{(P{\sep}Q)}
\\
\Leftrightarrow \GLOBAL{i}{0} {\ourimply} \hidefa{{\OP}}\code{(P}{\sep}\code{Q)}
\\
\Leftrightarrow \GLOBAL{i}{0} {\ourimply} \hidefa{\OP}\code{P} ~\sep~ \GLOBAL{i}{0} {\ourimply} \hidefa{\OP}\code{Q}
\\
\Leftrightarrow \LatchOut{i}{P}  {\sep} \LatchOut{i}{Q}
\end{array}
\]
\end{small}
\end{proof}

\begin{proof}[\splitrulen{2}]
\begin{small}
\[
\begin{array}{l}
\LatchIn{i}{(P{\sep}Q)}
\\
\Leftrightarrow {((\code{P} {\sep} \code{Q}) {\septract}\emp) {\sep} [\DEC]}_{\fracpermc} {\sep}\GLOBAL{i}{n} {\wedge}\code{n>0}
\\
\Leftrightarrow (\code{P} {\septract}\emp) {\sep} (\code{Q} {\septract}\emp) {\sep} [\DEC]_{\fracperm_1} {\sep}   {[\DEC]}_{\fracperm_2} {\sep}\GLOBAL{i}{n}{\sep}\GLOBAL{i}{n} {\wedge}\code{n>0}{\wedge}\fracpermc=\fracperm_1{+}\fracperm_2
\\
\Leftrightarrow (\code{P} {\septract}\emp) {\sep} [\DEC]_{\fracperm_1} {\sep}\GLOBAL{i}{n} {\wedge}\code{n>0} {\sep} (\code{Q} {\septract}\emp){\sep} [\DEC]_{\fracperm_2} {\sep}\GLOBAL{i}{n} {\wedge}\code{n>0}
\\
\Leftrightarrow \LatchIn{i}{P}  {\sep} \LatchIn{i}{Q}
\end{array}
\]
\end{small}
\end{proof}

\begin{proof}[\splitrulen{3}]
\begin{small}
\[
\begin{array}{l}
\CNT{i}{n} {\wedge} \code{n1,n2}{\geq}\code{0} {\wedge} \code{n{=}n1{+}n2}
\\
\Rightarrow \LOCAL{i}{n} {\wedge} \code{n1,n2}{\geq}\code{0} {\wedge} \code{n{=}n1{+}n2}
\\
\Rightarrow \LOCAL{i}{n1} {\sep} \LOCAL{i}{n2} {\wedge} \code{n1,n2}{\geq}\code{0} 
\\
\Rightarrow \CNT{i}{n1} {\sep} \CNT{i}{n2}
\end{array}
\]
\end{small}
\end{proof}

\begin{proof}[\normrulen{1}]
\begin{small}
\[
\begin{array}{l}
\CNT{c}{n} {\sep} \CNT{c}{-1} {\wedge} \code{n\sm{\leq}0}
\\
\Leftrightarrow  (\LOCAL{c}{n} {\wedge}\code{n}{\geq}\code{0} ~{\vee}~ \GLOBAL{c}{0}{\wedge}\code{n=-1}) {\sep} \CNT{c}{-1} {\wedge} \code{n\sm{\leq}0}
\\
\Leftrightarrow  (\LOCAL{c}{n} {\wedge}\code{n}{=}\code{0} ~{\vee}~ \GLOBAL{c}{0}{\wedge}\code{n=-1}) {\sep} \CNT{c}{-1}
\\
\Rightarrow  (\GLOBAL{c}{m} {\wedge}\code{m}{\geq}\code{0} ~{\vee}~ \GLOBAL{c}{0}{\wedge}\code{n=-1}) {\sep} \GLOBAL{c}{0}
\\
\Rightarrow \GLOBAL{c}{0}
\\
\Rightarrow \CNT{c}{-1}
\end{array}
\]
\end{small}
\end{proof}

\begin{proof}[\normrulen{2}]
\begin{small}
\[
\begin{array}{l}
\CNT{c}{n1} {\sep} \CNT{c}{n2}\, {\wedge}\, \code{n}{=}\code{n1{+}n2} \,{\wedge}\, \code{n1,n2}{\geq}\code{0}
\\
\Leftrightarrow \LOCAL{c}{n1} {\sep} \LOCAL{c}{n2}\, {\wedge}\, \code{n}{=}\code{n1{+}n2} \,{\wedge}\, \code{n1,n2}{\geq}\code{0}
\\
\Leftrightarrow \LOCAL{c}{n}\, {\wedge}\, \code{n}{=}\code{n1{+}n2} \,{\wedge}\, \code{n1,n2}{\geq}\code{0}
\\
\Rightarrow \CNT{c}{n}
\end{array}
\]
\end{small}
\end{proof}

\begin{proof}[\normrulen{3}]
\begin{small}
\[
\begin{array}{l}
\LatchOut{c}{P} {\sep} \CNT{c}{-1}
\\
\Leftrightarrow  (\GLOBAL{c}{0} {\ourimply} \code{P}) {\sep} \GLOBAL{c}{0} \\
\Rightarrow  \code{P} {\sep} \GLOBAL{c}{0} \\
\Rightarrow  \code{P} {\sep} \CNT{c}{-1} 
\end{array}
\]
\end{small}
\end{proof}

\begin{proof}[\errrulen{1}]
\begin{small}
\[
\begin{array}{l}
\LatchIn{c}{P}{\sep}\code{CNT(c,-1)}
\\
\Leftrightarrow {(\code{P} {\septract}\emp)  {\sep} [\DEC]_{\fracpermc} {\sep}\GLOBAL{c}{n} {\wedge}\code{n>0} {\sep}\code{CNT(c,-1)}}
\\
\Leftrightarrow {(\code{P} {\septract}\emp) {\sep} [\DEC]_{\fracpermc} {\sep}\GLOBAL{c}{n} {\wedge}\code{n>0} {\sep} \GLOBAL{c}{0} }
\\
\Leftrightarrow {(\code{P} {\septract}\emp) {\sep} ([\DEC]_{\fracpermc} {\sep}\GLOBAL{c}{n} {\wedge}\code{n>0} {\sep} \GLOBAL{c}{0}) }
\\
\Leftrightarrow ({\code{P} {\septract}\emp) {\sep} \false }
\\
\Leftrightarrow \false
\quad
//\RACE
\end{array}
\]
\end{small}
\end{proof}

\begin{proof}[\errrulen{2}]
\begin{small}
\[
\begin{array}{l}
\code{CNT(c,a)}{\sep}\code{CNT(c,-1)}\sm{\wedge}\code{a{>}0} 
\\
\Leftrightarrow \LOCAL{c}{a}{\sep}\code{CNT(c,-1)}\sm{\wedge}\code{a{>}0} 
\\
\Rightarrow \GLOBAL{c}{m}{\wedge} \code{m}{\geq}\code{a} {\sep} \GLOBAL{c}{0}\sm{\wedge}\code{a{>}0} 
\\
\Rightarrow \false
\quad
//\DEADLOCK
\end{array}
\]
\end{small}
\end{proof}

\hide{

\begin{verbatim}
Need to change:
  lc(i,-P) == DEC{f}*-P*[i->n]&n>0

Proofs for Splitting Lemmas
===========================
  // lc(i,-P*Q)
  // DEC{f}*-(P*Q)*[i->n]&n>0 & 0<f<=1
  // DEC{f1}*DEC{f2}*-P*-Q*[i->n]&&0<f1,f2<=1 & f=f1+f2 & n>0
  lc(i,-P*Q) --> lc(i,-P)*lc(i,-Q) 
  // DEC{f1}*-P*[i->n]*DEC{f2}*-Q*[i->n]&0<f1,f2<=1 & f=f1+f2 & n>0
  // lc(i,-P)*lc(i,-Q) 

  // lc(i,+P*Q)
  // +(P*Q)
  lc(i,+P*Q) --> lc(i,+P)*lc(i,+Q) 
  // P*Q)
  // lc(i,+P)*lc(i,+Q) 

  // CNT(i,n) & n1,n2>=0 & n=n1+n2 
  // {|i->n|} & n1,n2>=0 & n=n1+n2
  // {|i->n1|}*{|i->n2|} & n1,n2>=0 & n=n1+n2
  CNT(i,n) & n1,n2>=0 & n=n1+n2 --> CNT(i,n1)*CNT(i,n2)
  // {|i->n1|}*{|i->n2|} & n1,n2>=0 & n=n1+n2
  // CNT(i,n1)*CNT(i,n2) & n1,n2>=0 & n=n1+n2
  // CNT(i,n1)*CNT(i,n2)

Proofs for Normalization Lemmas
===============================
  // CNT(c,n) * CNT(c,-1) & n<=0 
  // {|c->n|}& n<=0 * {|c->-1|}
  // {|c->n|}& (n=-1|n=0) * [c->0]
  // ({|c->-1|} | {|c->0|})  * [c->0]
  // ([c->0] | [c->m]&m>=0)  * [c->0]
  CNT(c,n) * CNT(c,-1) & n<=0 ==> CNT(c,-1) // idempotent lemma 
  // ([c->0] | [c->0])
  // [c->0] 
  // CNT(c,-1) 


  // CNT(c,n1) * CNT(c,n2) & n=n1+n2 & n1,n2>=0 & n>=0  
  // {|c->n1|}*{|c->n2|} & n=n1+n2 & n1,n2>=0 & n>=0 
  // {|c->n1+n2|} & n=n1+n2 & n1,n2>=0 & n>=0 
  CNT(c,n1) * CNT(c,n2) & n=n1+n2 & n1,n2>=0 & n>=0  ==> CNT(c,n)
  // {|c->n|} & n=n1+n2 & n1,n2>=0 & n>=0 
  // CNT(c,n) & n=n1+n2 & n1,n2>=0 & n>=0 
  // CNT(c,n)

Proofs for Contradiction Lemmas
===========================================
  // lc(c,-P) * CNT(c,-1) & a>0 
  // DEC{f}* {|c->n>0|} * {|c->-1|} & a>0 
  // DEC{f}* [c->m] * [c->0] & m>0 
  lc(c,-P) * CNT(c,-1)   --> RACE-ERROR // race problem
  // [c->m /\ c->0] & m>0 
  // false
  // RACE-ERROR

  // CNT(c,a) * CNT(c,-1) & a>0 
  // {|c->a|} * {|c->-1|} & a>0 
  // [c->m] * [c->0] & m>=a>0 
  CNT(c,a) * CNT(c,-1) & a>0  --> DEADLOCK  // wait without any grant
  // [c->m /\ c->0] & m>=a>0 
  // false
  // DEADLOCK  
\end{verbatim}
}

The cycle detection lemma in \errrulen{3}
is a kind of contradiction detection mechanism
too, as explained in Sec~\ref{sec:deadlock}.
\hide{
To support multi-latch deadlocks, we have
also provided a \cc{WAIT(S)} set which is used to
capture waiting of the form \cc{c2{\ra}c1}, 
where \cc{c2} must complete before \cc{c1}.
In case we detect a cycle in such
wait-for arcs in \cc{S}, it would signal
a possible deadlock sceanario where a set of threads
are involved in some cyclic waiting.

However, if the waits-for set is cycle-free, we are
guaranteed to have some concurrency
schedule that would complete all the waiting
threads, based on the ordering of this waits-for
graph. This lemma cannot be proven as
a {\contra} lemma since it is not based
on some heap states. Instead, we have relied
on the expected ordering of execution to ensure
deadlock-freedom.
}

\section{Verification Proofs for a CDL Library}
\label{app:verified-methods}

For ease of presentation,
we have used this simple
version to illustrate how thread-local abstraction
are used.
Our verification tool can
automatically verify this
implementation of {\CDL}.
The detailed proof steps are re-produced below.
\wnnay{KILL is for destroy and is first converted to 
thread-local {|i->n|}}

\scnay{The  proof for {\createlatch} need patching, not sure how at the moment}

$
\begin{array}{l}
\code{{\CDL}\, {\createlatch}(int\, n) \code{with} \code{P} }
\\
\indent\quad \SPEC{\requires~\code{n\sm{\gt}0}} \\
\indent\quad \SPEC{\ensures~ \LatchIn{res}{P} {\sep}\LatchOut{res}{P}{\sep}\code{CNT(res,n)}} \\
\code{\{}
\\
\qquad \CSPEC{\code{n\sm{\gt}0}}
\\
\qquad
\code{{\CDL}\,i = new {\CDL}(n);}
\\
\qquad \CSPEC{i\sm{\mapsto}\CDL(n)\sm{\wedge}\code{n\sm{\gt}0}}
\\ 
\qquad \CSPEC{\sm{{[\DEC]}_{1} {\sep} {[\KILLR]}_{1} {\sep} \hidefa{\OM \code{P} {\sep}  \OP \code{P} {\sep}} \GLOBAL{i}{n}}\sm{\wedge}\code{n\sm{\gt}0}}
\\ 
\qquad \CSPEC{\sm{{[\DEC]}_{1} {\sep} \LOCAL{i}{n} {\sep} \GLOBAL{i}{n}}\sm{\wedge}\code{n\sm{\gt}0}}
\\ 
\qquad \CSPEC{\sm{{[\DEC]}_{1} {\sep} \LOCAL{i}{n} {\sep} (\code{P}{\septract}\emp){\sep} \code{P} {\sep} \GLOBAL{i}{n}}\sm{\wedge}\code{n\sm{\gt}0}}
\\ 
\qquad \CSPEC{ (\code{P}{\septract}\emp){\sep} \sm{{[\DEC]}_{1}   {\sep} \GLOBAL{i}{n} {\sep} (\GLOBAL{i}{0} {\ourimply} \code{P}) {\sep} \LOCAL{i}{n} {\wedge}\code{n\sm{\gt}0}} }
\\
\qquad \CSPEC{\sm{\LatchIn{i}{P} {\sep} \LatchOut{i}{P} {\sep} \CNT{i}{n}}}
\\
\qquad \code{return i; }
\\
\qquad \CSPEC{\sm{\LatchIn{\res}{P} {\sep} \LatchOut{\res}{P} {\sep} \CNT{\res}{n}}}
\\
\code{\}}
\end{array}
$
\\[1ex]
\indent
$
\begin{array}{l}
\code{{\CDL}\, {\createlatch}(int\, n) \code{with} \code{P} }
\\
\indent\quad \SPEC{\requires~\code{n\sm{=}0}} \\
\indent\quad \SPEC{\ensures~ \code{CNT(res,-1)}} \\
\code{\{}
\\
\qquad \CSPEC{\code{n\sm{=}0}}
\\
\qquad
\code{{\CDL}\,i = new {\CDL}(n);}
\\
\qquad \CSPEC{i\sm{\mapsto}\CDL(n)\sm{\wedge}\code{n\sm{=}0}}
\\ 
\qquad \CSPEC{\GLOBAL{i}{n}\sm{\wedge}\code{n\sm{=}0}}
\\
\qquad \CSPEC{\CNT{i}{-1}}
\\
\qquad\code{return i; }
\\
\qquad \CSPEC{\CNT{res}{-1}}
\\
\code{\}}
\end{array}
$
\\[1ex]
\indent
$
\begin{array}{l}
\code{void~countDown(\CDL~i)}\\
\indent\quad \SPEC{\requires~\LatchIn{i}{P}{\sep}\code{P}{\sep}\code{CNT(i,n)}\sm{\wedge}\code{n{>}0}}\\
\indent\quad \SPEC{\ensures~ \code{CNT(i,n{-}1)}};\\ 
\code{\{}
\\
\qquad \CSPEC{\LatchIn{i}{P}{\sep}\code{P}{\sep}\code{CNT(i,n)}\sm{\wedge}\code{n{>}0}}
\\
\qquad \CSPEC{\sm{( \code{P} {\septract}\emp){\sep} {[\DEC]}_{\fracpermc} {\sep} \GLOBAL{i}{m} {\wedge}\code{m{>}0}} {\sep}~\code{P}~{\sep} \LOCAL{i}{n} \sm{\wedge}\code{n\sm{>}0}}
\\
\qquad \CSPEC{\sm{{[\DEC]}_{\fracpermc} {\sep}\GLOBAL{i}{m} {\wedge}\code{m{>}0}} {\sep} \LOCAL{i}{n} \sm{\wedge}\code{n\sm{>}0}}
\\
\qquad \code{\criticalSec{if (i.val>0) i.val=i.val-1;}}
\\
\qquad \CSPEC{\sm{\GLOBAL{i}{m{-}1} {\wedge}\code{m{>}0} } {\sep} \LOCAL{i}{n{-}1} \sm{\wedge}\code{n\sm{>}0}}
\\
\qquad \CSPEC{\LOCAL{i}{n{-}1} \sm{\wedge}\code{n\sm{>}0}}
\\
\qquad \CSPEC{\CNT{i}{n{-}1}}
\\
\code{\}}
\end{array}
$
\\[1ex]
\indent
$
\begin{array}{l}
\code{void~countDown(\CDL~i)}\\
\indent\quad \SPEC{\requires~\code{CNT(i,-1)}}\\
\indent\quad \SPEC{\ensures~ \code{CNT(i,-1)}};\\
\code{\{}
\\
\qquad \CSPEC{\code{CNT(i,-1)}}
\\
\qquad \CSPEC{\GLOBAL{i}{0}}
\\
\qquad\code{\criticalSec{if (i.val>0) i.val=i.val-1;}}
\\
\qquad \CSPEC{\GLOBAL{i}{0}}
\\
\qquad \CSPEC{\code{CNT(i,-1)}}
\\
\code{\}}
\end{array}
$
\\[1ex]
\indent
$
\begin{array}{l}
\code{void~await(\CDL~i)}\\
\indent\quad \SPEC{\requires~\LatchOut{i}{P}{\sep}\code{CNT(i,0)}}\\
\indent\quad \SPEC{\ensures~ \code{P}{\sep}\code{CNT(i,-1)}};\\
\code{\{}
\\
\qquad \CSPEC{\LatchOut{i}{P}{\sep}\code{CNT(i,0)}}
\\
\qquad \CSPEC{(\GLOBAL{i}{0}{\ourimply}\code{P}){\sep}\LOCAL{i}{0}}
\\
\qquad \CSPEC{(\GLOBAL{i}{0}{\ourimply}\code{P}){\sep}\GLOBAL{i}{m}\sm{\wedge}\code{m\sm{\geq}0}}
\\
\qquad \code{while (i.val>0) skip; }
\\
\qquad \CSPEC{(\GLOBAL{i}{0}{\ourimply}\code{P}){\sep}\GLOBAL{i}{0} }
\\
\qquad \CSPEC{\code{P}{\sep}\code{CNT(i,-1)}}
\\
\code{\}}
\end{array}
$

\indent
$
\begin{array}{l}
\code{void~await(\CDL~i)}\\
\indent\quad \SPEC{\requires~\code{CNT(i,-1)}}\\
\indent\quad \SPEC{\ensures~\code{CNT(i,-1)}};\\
\code{\{}
\\
\qquad \CSPEC{\code{CNT(i,-1)}}
\\
\qquad \CSPEC{\GLOBAL{i}{0}}
\\
\qquad \code{while (i.val>0) skip; }
\\
\qquad \CSPEC{\GLOBAL{i}{0}}
\\
\qquad \CSPEC{\code{CNT(i,-1)}}
\\
\code{\}}
\end{array}
$

\hide{
\begin{verbatim}
  CDL newCDL(n) with P
   requires n>=0  // exception thrown when n is negative
   ensures lc(res,+P) * lc(res,-P) * CNT(res,n)
  {
    // emp
    i=alloc(1)
    // i->_
    i.val=n;
    // i->n
    // DEC * P * -P * [i->n] & n>=0
    // DEC * P * -P * {|i->n|}
    // lc(i,+P)*lc(i,-P)*CNT(i,n)
    return i;
    // lc(res,+P)*lc(res,-P)*CNT(res,n)
  }

 void countDown(CDL i) 
   requires lc(i,-P)*P*CNT(i,n) & n>0
   ensures  CNT(i,n-1);
   requires CNT(i,-1) 
   ensures  CNT(i,-1);
 {
   // lc(i,-P)* P * CNT(i,n) & n>0
   // [i->c] & c>0 * -P * P * {|i->n|} & n>0 
   // [i->c] & c>0 * {|i->n|} & n>0
   // [i->c] & c>0 * {|i->n|} & n>0
    <if i.val>0 then i.val=i.val-1>
   // [i->c-1] & c>0 * {|i->n-1|} & n>0
   // {|i->n-1|} & n>0
   // CNT(i,n-1) & n>0
   // CNT(i,n-1) 
 }
   requires CNT(i,-1) 
   ensures  CNT(i,-1);
 {
   // CNT(i,-1) 
   // {|i->-1|} 
   // [i->0] 
    <if i.val>0 then i.val=i.val-1>
   // [i->0] 
   // {|i->-1|}
   // CNT(i,-1)
 }

  void await(i)
   requires lc(i,+P) * CNT(i,n) & n<=0
   ensures  P * CNT(i,-1);
   {
   // lc(i,+P) * CNT(i,n) & n<=0
   // P *{|i->n|} & n<=0
   // P *{|i->n|} & (n=-1  | n=0)
   // P * ([i->0] | [i->m] & m>=0 & n=0)
   while(i.val>0)
     skip; 
   // P * [i->0] 
   // P * {(|i->-1|} 
   // P*CNT(i,-1)
  }
\end{verbatim}
}









\end{document}